\documentclass[aps,prd,preprintnumbers,showpacs,nofootinbib]{revtex4}

\usepackage{overpic}
\usepackage{subfigure}
\usepackage[english]{babel}
\usepackage{amsmath}
\usepackage{amssymb}
\usepackage{epsfig}
\usepackage{graphics,psfrag,rotating}
\usepackage{graphicx}
\usepackage{dcolumn}
\usepackage{bm}
\begin{document}
\newcommand{\Od}{{\cal O}}
\newcommand{\lsim}   {\mathrel{\mathop{\kern 0pt \rlap
  {\raise.2ex\hbox{$<$}}}
  \lower.9ex\hbox{\kern-.190em $\sim$}}}
\newcommand{\gsim}   {\mathrel{\mathop{\kern 0pt \rlap
  {\raise.2ex\hbox{$>$}}}
  \lower.9ex\hbox{\kern-.190em $\sim$}}}


\title{Black Holes  in $f(R)$ theories}%

\author{A. de la Cruz-Dombriz\footnote{E-mail: dombriz@fis.ucm.es},
A. Dobado\footnote{E-mail: dobado@fis.ucm.es}%
 \,\,and A. L. Maroto\footnote{E-mail: maroto@fis.ucm.es}%
}
\affiliation{Departamento de  F\'{\i}sica Te\'orica I, Universidad Complutense
 de Madrid, 28040 Madrid, Spain.
}
\date{\today}

\begin{abstract}
In the context of $f(R)$ theories of gravity, we address the problem of finding
static and spherically symmetric black hole solutions. Several aspects
of  constant curvature solutions with and without electric charge are discussed. We also
study the general case (without imposing constant curvature). Following a perturbative
approach around the Einstein-Hilbert action, it is found that only solutions of the
Schwarzschild-(Anti) de Sitter type are present up to second order
in perturbations. Explicit expressions for the effective cosmological constant
are obtained in terms of the $f(R)$ function.
Finally, we have considered the thermodynamics of black holes in Anti-de Sitter
space-time and found that this kind of solutions can only exist provided the
theory satisfies $R_0+f(R_0)<0$. Interestingly, this expression is related to
the condition which guarantees the positivity of the effective Newton's constant
in this type of theories.  In addition, it also ensures that the thermodynamical properties
in $f(R)$ gravities are qualitatively similar to those of
standard General Relativity.
\end{abstract}

\maketitle

\section{Introduction}

In the last years, increasing attention has been paid to modified theories of gravity
in order  to understand several open cosmological questions such as the accelerated
expansion of the universe \cite{Carroll}
and the dark matter origin \cite{Cembranos:2008gj}.
Some of those theories modify
General Relativity by adding higher powers of the scalar curvature $R$, the
Riemann and Ricci tensors or their derivatives \cite{Maroto&Dobado_1993}.
Lovelock and $f(R)$
theories are some examples of these attempts. It is therefore quite natural to ask
about black holes (BH) features in those gravitational theories since, on the one
hand, some BH signatures may be peculiar to Einstein's gravity and others may be
robust features of all generally covariant theories of gravity. On the other hand,
the results obtained may lead to rule out some models which will be in desagreement with
expected physical results. For thoses purposes, research on thermodynamical quantities
of BH is of particular interest.

In this work we will restrict ourselves to the so called $f(R)$ gravity
theories (see \cite{Sot}) in metric formalism in Jordan's frame. In this frame, the gravitational
Lagrangian
is given by $R+f(R)$  where $f(R)$ is an arbitrary function of $R$ and Einstein's
equations are
usually fourth order in the metric (see \cite{varia} for several proposed $f(R)$
functions compatible with local gravity tests and other cosmological constraints).
An alternative approach would be to use the
Einstein's frame, where ordinary Einstein's gravity coupled to a scalar plus a massive
spin-2 field is recovered. Even if a mathematical correspondence could be established
between those two frames, in the last years some controversy has remained about their
physical equivalence.

Previous literature on $f(R)$ theories \cite{Whitt}  proved in Einstein's frame
that
Schwarzschild solution is the only  static spherically
symmetric solution for an action  of the form $R+aR^2$ in $D=4$.
In \cite{Mignemi} uniqueness theorems of spherically symmetric solutions for
general polynomial actions in arbitrary dimensions using Einstein's frame were proposed
(see also \cite{Multamaki} for additional results). See also \cite{olmo} for
spherical solutions with sources.

Using the euclidean action method (see for instance \cite{Hawking&Page, Witten})
in order to determine different thermodynamical quantities,
Anti de Sitter ($AdS$) BH  in $f(R)$ models have been
studied \cite{Cognola}. In  \cite{Briscese} the entropy of
Schwarzschild-de Sitter  BH
 was calculated for some particular cosmologically
viable models in vacuum and their cosmological stability was discussed.

BH properties have been also widely studied in other modified gravity theories.
For instance, \cite{cvetic,Cai_GaussBonet_AdS} studied  BH
 in Einstein's theory  with a Gauss-Bonnet
term and  cosmological constant. Different results were found depending on
the dimension $D$ and the sign of the constant horizon curvature $k$.
For $k=0,-1$, the Gauss-Bonnet term does not
modify $AdS$ BH thermodynamics at all (only the horizon position is modified with
respect to the Einstein-Hilbert ($EH$) theory) and BH are not only
locally thermodynamically stable but
also globally preferred. Nevertheless for $k=+1$ and $D=5$ (for $D\geq6$ thermodynamics
is again essentially that for $AdS$ BH) there exist some features not present in
the absence of Gauss-Bonnet term. Gauss-Bonnet and/or Riemann squared interaction terms
were studied in \cite{Cho} concluding that in this case phase transitions may occur
with $k=-1$ .

Another approach is given by Lovelock gravities, which are free of ghosts and where
field equations contain no more than second derivatives of the metric. These theories
were studied in \cite{Matyjasek} and the corresponding entropy was evaluated.

The paper is organized as follows: in section 2 we present some
general results for $f(R)$
gravities for interesing physical situations in metric formalism.
In sections 3 and 4, BH in $f(R)$
gravities are studied and explicit Einstein's field equations are presented for
static and spherically symmetric metrics. Section 5 is devoted to find
perturbative solutions for static and spherically symmetric background metric:
general metric coefficients are found depending on $f(R)$ derivatives evaluated
at background scalar curvature. Sections 6 and 7 are widely devoted to study
thermodynamical quantities and their consequences in local and global stability
for some particular $f(R)$ models. Finally, we include some conclusions.

\section{General Results}

In order to study the basics of the solutions of general $f(R)$ gravity theories,
let us start from the action
\begin{equation}
S\,=\,S_g+S_m
\end{equation}
where $S_g$ is the $D$ dimensional gravitational action:
\begin{equation}
S_g=\frac{1}{16 \pi G_D}\int \text{d}^{D}x\sqrt{\mid g\mid}\,(R+f(R))
\label{S_g}
\end{equation}
with $G_D \equiv M_D^{2-D}$ being the $D$ dimensional Newton's
constant,  $M_D$ the corresponding Planck mass, $g$ the determinant
of the metric $g_{AB}$, $(A,B=0, 1, ..., D-1)$, $R$ the scalar
curvature and $R+f(R)$ is the function defining the theory under consideration.
As the simplest example, the $EH$ action with cosmological
constant $\Lambda_D$ is given by $f(R)=-(D-2)\Lambda_D$.
\\
The matter action $S_m$ defines the energy-momentum tensor as:
\begin{equation}
T^{AB}=-\frac{2}{\sqrt{\mid g\mid}}\frac{\delta S_m}{\delta
g_{AB}}.
\end{equation}
From the above action, the equations of motion in the metric formalism are just:
\begin{eqnarray}
 R_{AB}(1+f'(R)) - \frac{1}{2}(R+f(R))\,g_{AB}+
(\nabla_A \nabla_B-g_{AB}\Box)f'(R)+8\pi G_D T_{AB}=0
\label{Einsteins_eqns}
\end{eqnarray}
where $R_{AB}$ is as usual the Ricci tensor and $\Box=\nabla_A \nabla^A$ with $\nabla$
the usual covariant derivative.
Thus for the vacuum $EH$ action with cosmological constant we have:
\begin{eqnarray}
 R_{AB}-\frac{1}{2}R\,g_{AB}+\frac{D-2}{2}\Lambda_D g_{AB}=0
\end{eqnarray}
which means  $R_{AB}=\Lambda_D g_{AB}$ and $R= D \Lambda_D$.
Coming back to the general case, the required condition to get
constant scalar curvature solutions $R\,=\,R_0$ (from now $R_0$
will denote a constant curvature value) in vacuum  implies:
\begin{eqnarray}
 R_{AB}\,(1+f'(R))-\frac{1}{2}\,g_{AB}\,(R+f(R))\,=\,0
\end{eqnarray}
Taking the trace in previous equation, $R_0$ must be a root of the equation:
\begin{eqnarray}
 2(1+f'(R_0))\,R_{0}-D\,(R_{0}+f(R_{0}))\,=\,0
\label{root_R0}
\end{eqnarray}
For this kind of solution an effective cosmological constant may be defined
 as $\Lambda_D^{eff}\equiv R_{0}/D$. Thus any constant curvature
solution $R=R_0$ with $1+f'(R_0)\neq 0$ fulfills:
\begin{eqnarray}
 R_{AB}=\frac{R_{0}+f(R_0)}{2(1+f'(R_0))}\,g_{AB}
\end{eqnarray}
On the other hand one can consider:
\begin{equation}
2R\,(1+f'(R))-D\,(R+f(R))\,=\,0
\label{dif}
\end{equation}
as a differential equation for the $f(R)$ function so that the corresponding
solution would admit any curvature $R$ value. The
solution of this differential equation is just:
\begin{equation}
f(R)\,=\,a R^{D/2}-R
\end{equation}
where $a$ is an arbitrary constant. Thus the gravitational  Lagrangian  becomes
 proportional to $a R^{D/2}$ which will have solutions of constant curvature for arbitrary $R$. The
reason is that this action is scale invariant since $a/G_D$ is
a non-dimensional constant.

Now we will address the issue of finding some general criteria to relate solutions of the $EH$
 action with solutions of more general $f(R)$ gravities, not necessarily of
 constant curvature $R$. Let $g_{AB}$ a solution of $EH$ gravity with
cosmological constant, i.e.:
\begin{eqnarray}
 R_{AB}-\frac{1}{2}R\,g_{AB}+\frac{D-2}{2}\Lambda_D g_{AB}+8\pi G_D
 T_{AB}=0
\label{LambdaCDM_eq}
\end{eqnarray}
Then $g_{AB}$ is also a solution of any $f(R)$ gravity, provided the following
compatibility equation
\begin{eqnarray}
f'(R) R_{AB}-\frac{1}{2}g_{AB}\left[f(R)+(D-2)\Lambda_D\right]
+(\nabla_A\,\nabla_B-g_{AB}\Box)f'(R)=0
\label{comp}
\end{eqnarray}
obtained from (\ref{Einsteins_eqns}) is fulfilled.
In the following we will consider  some particularly
interesting cases. The simplest possibility is obviously vacuum
($T_{AB}=0$) with vanishing cosmological constant $\Lambda_{D}=0$.
Then the above equation \eqref{LambdaCDM_eq} becomes:
\begin{equation}
R_{AB}=\frac{1}{2}R g_{AB}
\end{equation}
which implies $R=0$ and $R_{AB}=0$. Consequently
 $g_{AB}$ is also a solution of any $f(R)$ gravity provided $f(0)=0$, which is for instance the case
when $f(R)$ is analytical around $R=0$. When the cosmological constant is different from zero
($\Lambda_{D}\neq 0$), but still  $T_{AB}=0$, we have also constant curvature with
$R_0=D\Lambda_D$ and $R_{AB}=\Lambda_D g_{AB}$. Then the
compatibility equation (\ref{comp}) reduces to (\ref{root_R0}).
In other words, $g_{AB}$ is also a solution of $f(R)$
provided $f(D\Lambda_D)=\Lambda_D(2-D+2f'(D\Lambda_D))$.
Notice that it would also be a solution for any $R_0$ in the particular case
$f(R)=aR^{D/2}-R$.
\\

Next we can consider the case with $\Lambda_D = 0$
and conformal matter ($T=T_A^A=0$).
 For a perfect fluid this means having the  equation of state
$\rho=(D-1)p$ where $p$ is the pressure and $\rho$ the energy density. In this case \eqref{LambdaCDM_eq} implies
\begin{eqnarray}
R\,=\,0\,\,\,;\,\,\, R_{AB}\,=\,8\pi G_D T_{AB}
\label{eq_conformal_matter}
\end{eqnarray}
Then, provided $f(0)=f'(0)=0$, $g_{AB}$ is also a solution of any
$f(R)$ gravity. This result could have particular interest in
cosmological  calculations for ultrarelativistic matter (i.e.
conformal) dominated universes. For the case of conformal matter with non vanishing $\Lambda_D$ we have again
constant $R=R_0$ with $R_0=D \Lambda_{D}$ and $g_{AB}$ is a solution of $f(R)$
provided that once again $f(D\Lambda_D)=\Lambda_D(2-D+2f'(D\Lambda_D))$.

\section{Black Holes in $f(R)$ gravities}

Now we consider the external metric for the gravitational field
produced by  a non rotating object in $f(R)$ gravity theories. The
most general static and spherically symmetric $D\geq 4$
dimensional metric can be written as (see \cite{ortin}):
\begin{eqnarray}
\text{d}s^2\,=\,e^{-2\Phi(r)} A(r)\text{d}t^2-A^{-1}(r)\text{d}
r^2-r^2\text{d}\Omega_{D-2}^2
\label{metric_D_v1}
\end{eqnarray}
or alternatively
\begin{eqnarray}
\text{d}s^2\,=\,\lambda(r)\text{d}t^2-\mu^{-1}(r)\text{d}r^{2}-r^2\text{d}\Omega_{D-2}^2
\label{metric_D_v2}
\end{eqnarray}
where $\text{d}\Omega_{D-2}^2$ is the metric on the $S^{D-2}$ sphere and identification
$\lambda(r)=e^{-2\Phi(r)}A(r)$ and $\mu(r)=A(r)$ can be straightforwardly established.

For obvious reasons the $\Phi(r)$ function is called the anomalous
redshift. Notice that a photon emitted at $r$ with proper
frequency $\omega_0$ is measured at infinity with frequency
$\omega_{\infty}= e^{-\Phi(r)}\sqrt{A(r)}\omega_0$.
%
As the metric is static, the scalar curvature $R$ in $D$ dimensions depends only on  $r$ and
it is given, for the metric parametrization \eqref{metric_D_v1}, by:
\begin{eqnarray}
R(r)\,  &=& \,\frac{1}{r^2}[D^2-5 D+6+r A'(r) \left(-2 D+3 r
\Phi '(r)+4\right)\nonumber\\&-&r^2 A''(r)-A(r) \left(D^2-5 D+2
r^2 \Phi '(r)^2-2 (D-2) r \Phi '(r)-2 r^2 \Phi ''(r)+6\right)].
\label{Dcurv}
\end{eqnarray}
where the prime denotes derivative with respect to $r$.
At this stage it is interesting to ask about which are the most
general static and spherically symmetric metrics with constant
scalar curvature $R_{0}$. This curvature can be found solving the
equation $R=R_0$. Then it is inmediate  to see that for a constant $\Phi(r)=\Phi_{0}$  the
general solution is:
\begin{eqnarray}
A(r)\,=\,1+a_{1}r^{3-D}+a_{2}r^{2-D}-\frac{R_0}{D(D-1)}r^2
\label{A_solution_R_constant_Dobado_procedure}
\end{eqnarray}
with $a_{1}$ and $a_{2}$ being arbitrary integration constants. In fact,
for the particular case $D=4$, $R_{0}=0$ and $\Phi_{0}=0$, the
metric can be written exclusively in terms of the function:
\begin{eqnarray}
A(r)\,=\,1+\frac{a_{1}}{r}+\frac{a_{2}}{r^{2}}.
\label{RN_solution_R_constant_Dobado_procedure}
\end{eqnarray}
By establishing the identifications $a_{1}=-2G_{N}M$ and
$a_{2}=Q^2$, this solution corresponds to a Reissner-Nordstr\"{o}m solution, ie. a charged massive BH solution with mass $M$
and charge $Q$. Further comments about this result will be made
below.

\section{Constant curvature black-hole solutions}

By inserting the metric \eqref{metric_D_v1} into the general
$f(R)$ gravitational action $S_g$ in (\ref{S_g}), and making variations with
respect to the $A(r)$ and $\Phi(r)$ functions, we find the
equations of motion:
\begin{eqnarray}
(2-D ) (1+f'(R)) \Phi'(r)-r\left[f'''(R)
R'(r)^2+f''(R)(\Phi'(r)R'(r)+ R''(r))\right]\,=\,0
\label{eqn_A}
\end{eqnarray}
and
\begin{eqnarray}
&& 2 r A(r) f'''(R) R'(r)^2+ f''(R)[2 D A(r)R'(r)
- 4 A(r) R'(r) + 2 r A(r) R''(r)+A'(r) r R'(r)]+\nonumber\\
&&g'(R)[-2 r A(r) \Phi'(r)^{2} + 2 D A(r) \Phi'(r) -4 A(r)
\Phi'(r) - r A''(r)  + 2 r A(r) \Phi''(r)+    \nonumber\\
&&         A'(r)(2 - D  + 3 r \Phi'(r)) ] - r(R+f(R))\,=\,0
\label{eqn_phi}
\end{eqnarray}
where $f'$, $f''$ and $f'''$ denote derivatives of $f(R)$ with
respect to the curvature $R$. \\
The above equations look in principle quite difficult to solve. For this reason
we will firstly consider the case of constant scalar curvature
 $R=R_0$ solutions. Then the equations of motion reduce to:
\begin{equation}
(2-D)\,(1+f'(R))\Phi'(r)=0
\label{eq_motion_phi}
\end{equation}
and
\begin{equation}
R+f(R)+(1+f'(R))\left[A''(r)+(D-2)\frac{A'(r)}{r}
-(2D-4)\frac{A(r)\Phi'(r)}{r}-3A'(r)\Phi'(r)+2A(r)\Phi'^2(r)-2A(r)\Phi''(r)\right]\,=\,0
\label{eq_motion_A}
\end{equation}
As commented in the previous sections, the constant curvature solutions of
$f(R)$ gravities are given by:
\begin{equation}
R_0=\frac{D\,f(R_0)}{2(1+f'(R_0))-D}
\label{const}
\end{equation}
whenever $2(1+f'(R_0))\neq D$. Thus from \eqref{eq_motion_phi}  $\Phi'(r)=0$ and then \eqref{eq_motion_A} becomes
\begin{equation}
R_{0}+f(R_0)+(1+f'(R_0))\left[A''(r)+(D-2)\frac{A'(r)}{r}\right]\,=\,0
\label{eqn_A_determination}
\end{equation}

Coming back to \eqref{eqn_A_determination}, and using \eqref{const}, we get
\begin{equation}
A''(r)+(D-2)\frac{A'(r)}{r}=-\frac{2}{D}R_0
\label{A_eq_R_constant}
\end{equation}
This is a $f(R)$-independent linear second order inhomogeneous differential
equation which can be easily integrated to give the general solution:
\begin{equation}
A(r)\,=\,C_1\,+\,C_2r^{3-D}-\frac{R_0}{D(D-1)}r^2
\label{A_solution_R_constant}
\end{equation}
which depends on two arbitrary constants $C_1$ and $C_2$. However
this solution has no constant curvature in
 the general case since, as we found above, the constant curvature requirement
demands $C_{1}=1$. Then, for negative $R_0$, this solution
  is basically the $D$ dimensional generalization obtained by Witten \cite{Witten}
of the BH in $AdS$ space-time
   solution considered by Hawking and Page \cite{Hawking&Page}. With
the natural choice $\Phi_0=0$ the
solution can be written as:
\begin{equation}
A(r)=1-\frac{R_{S}^{D-3}}{r^{D-3}}+\frac{r^2}{l^2}.
\end{equation}
where
\begin{equation}
R_S^{D-3}=\frac{16\pi G_D M}{(D-2)\mu_{D-2}} \label{BHmass}
\end{equation}
with
\begin{equation}
\mu_{D-2}=\frac{2\pi^{\frac{D-1}{2}}}{\Gamma(\frac{D-1}{2})}
\end{equation}
being the area of the $D-2$ sphere, $l^2\equiv-D(D-1)/R_0 $ is the asymptotic $AdS$ space scale squared and $M$ is the mass parameter usually found in the literature.

Thus we have concluded  that the only static and
spherically symmetric vacuum solutions with constant (negative) curvature of any
$f(R)$ gravity is just the Hawking-Page BH in $AdS$ space.
However this kind of solution is not the most general static and
spherically symmetric metric with constant curvature as can be seen by
comparison with the solutions found in
(\ref{A_solution_R_constant_Dobado_procedure}).
Therefore we have to conclude that there are constant curvature
BH solutions that cannot be obtained as vaccum solutions of any $f(R)$
theory. As we show below, in the $D=4$ case,  we see that
the most general case can be described as
a charged BH solution in $f(R)$-Maxwell theory.

Indeed, let us consider now the case of charged black holes in $f(R)$ theories.
We will limit ourselves to  the $D=4$ case, since in other dimensions
the curvature is not necessarily constant. The action of the theory
is now the generalization of the Einstein-Maxwell action:
\begin{equation}
S_g=\frac{1}{16 \pi G_4}\int \text{d}^{4}x\sqrt{\mid g\mid}\,(R+f(R)-F_{\mu\nu}F^{\mu\nu})
\end{equation}
where $F_{\mu\nu}=\partial_\mu A_\nu-\partial_\nu A_\mu$.
Considering an electromagnetic potential of the form: $A_\mu=(V(r),\vec 0)$ and the  static
spherically symmetric metric (\ref{metric_D_v1}), we find that the solution with
constant curvature $R_0$ reads:
\begin{eqnarray}
V(r)&=&\frac{Q}{r}\nonumber \\
\lambda(r)&=&\mu(r)=1-\frac{2G_{4}M}{r}+\frac{(1+f'(R_0))Q^2}{r^2}-\frac{R_0}{12}r^2
\end{eqnarray}
Notice that unlike the $EH$ case, the contribution of the black-hole charge
to the metric tensor is corrected by a $(1+f'(R_0))$ factor.

\section{Perturbative results}

In the previous section we have considered static spherically symmetric
solutions with  constant curvature. In $EH$ theory this would provide the most general
solution with spherical symmetry. However, it is not guaranteed this to be the case
also in $f(R)$ theories. The problem of finding the general  static spherically
symmetric solution in arbitrary $f(R)$ theories without imposing the
constant curvature condition is in principle too complicated. For that reason in this section
we will present a perturbative analysis of the problem, assuming that
the modified action is a small perturbation around $EH$ theory.

Therefore  let us consider a $f(R)$ function  of the form
\begin{eqnarray}
f(R)\,=-(D-2)\Lambda_{D}+\alpha g(R)
\label{expansion_en_alpha_fR}
\end{eqnarray}
where $\alpha\ll 1$ is a dimensionless parameter and $g(R)$ is assumed
to be analytic in $\alpha$. By using the metric parametrization given by
\eqref{metric_D_v2} the equations of motion become:
\begin{eqnarray}
\lambda (r) (1&+&f'(R)) \left\{2 \mu (r) \left[(D -2) \lambda '(r)
+r \lambda ''(r)\right]+r \lambda '(r) \mu '(r)\right\}\nonumber \\
&-&2 \lambda (r)^2 \left\{2 \mu(r)[(D -2) R'(r) f''(R)
+r f^{(3)}(R) R'(r)^2+r R''(r) f''(R)]+r R'(r) \mu '(r) f''(R)\right\}\nonumber\\
&-&r \mu (r) \lambda '(r)^2 (1+f'(R))+2 r \lambda (r)^2(R+f(R))=\,0
\label{eqn_lambda}
\end{eqnarray}

\begin{eqnarray}
&-&\lambda (r) \mu '(r) \left[2 (D -2) \lambda (r)
+r \lambda '(r)\right] (1+f'(R))\nonumber \\
&+&\mu (r) \Big\{2 \lambda (r) R'(r)
\left[2 (D -2) \lambda(r)+r\lambda '(r)\right] f''(R)
+r(1+f'(R))(\lambda '(r)^2-2 \lambda (r) \lambda ''(r))\Big\}\nonumber \\
&-&2 r \lambda (r)^2 (R+f(R))\,=\,0
\label{eqn_mu}
\end{eqnarray}
where prime denotes derivative with respect to
the corresponding argument and $R\equiv R(r)$ is given by (\ref{Dcurv}).
Now, assuming that the $\lambda(r)$ and
$\mu(r)$ functions appearing in the metric \eqref{metric_D_v2} are also
analytical in $\alpha$, they can
be written as follows
\begin{eqnarray}
\lambda(r)\,&=&\,\lambda_{0}(r)+\sum_{i=1}^{\infty}\alpha^{i}\lambda_{i}(r)    \nonumber\\
\mu(r)\,&=&\,\mu_{0}(r)+\sum_{i=1}^{\infty}\alpha^{i}\mu_{i}(r)
\label{expansion_en_alpha_lambda&mu}
\end{eqnarray}
where $\{\lambda_{0}(r),\,\mu_{0}(r)\}$ are the unperturbed solutions for the
$EH$ action with cosmological constant given by
\begin{eqnarray}
\mu_{0}(r)\,&=&\,1+\frac{C_1}{r^{D-3}}-\frac{\Lambda_{D}}{(D-1)}r^2\nonumber\\
\lambda_{0}(r)\,&=&\,-C_{2}(D-2)(D-1)\,\mu_{0}(r)
\label{mu0_lambda0}
\end{eqnarray}
which are the standard BH solutions in a $D$ dimensional $AdS$
spacetime. Note that the factor $C_2$ can be chosen by
performing a coordinate $t$ reparametrization so that both
functions could be indentified. For the moment, we will keep the
background solutions as given in \eqref{mu0_lambda0} and we will
discuss the possibility of getting $\lambda(r)=\mu(r)$  in the
 perturbative expansion later on.

By inserting  \eqref{expansion_en_alpha_fR} and
\eqref{expansion_en_alpha_lambda&mu} in \eqref{eqn_lambda} and
\eqref{eqn_mu} we obtain the following first order equations:
\begin{eqnarray}
(D-3)\mu_{1}(r)+r\mu_{1}'(r)+\frac{2\Lambda_{D}g'(R_0)-g(R_0)}{D-2}r^{2}\,=\,0
\label{lambda_eqn1}
\end{eqnarray}
\begin{eqnarray}
&C_{2}& \left[C_{1} (D-1) r^{3-D}-\Lambda_{D} r^2+D-1\right]g(R_{0}) r^2+\left[C_{1} (D-3)
 r^{3-D}+\frac{2\Lambda_{D}}{D-1}r^{2}\right]\lambda_{1}(r)\nonumber\\
&+&C_{2} (D-2)(D-1)\left(\Lambda_{D} r^2-D+3\right)\mu_{1}(r)\nonumber\\
&+&\left(1+C_{1} r^{3-D}-\frac{\Lambda_{D} r^2}{D-1}\right)
 \left[2 C_{2} (1-D) r^2\Lambda_{D}g'(R_0)+r\lambda_{1}'(r)\right]\,=\,0\nonumber\\
&&
\label{mu_eqn1}
\end{eqnarray}
whose solutions are:
\begin{eqnarray}
\lambda_{1}(r)\,&=&\,C_{4}(D-1)(D-2)+\frac{(C_{1}C_{4}-C_{2}C_{3})(D-2)(D-1)}{r^{D-3}}
\nonumber \\
&-&\left[C_{4}(D -2)
\Lambda_{D}+C_{2}\left(g(R_0)-2\Lambda_{D} g'(R_{0})\right)\right] r^{2}
\nonumber\\
&&
\label{mu1}
\end{eqnarray}
\begin{eqnarray}
\mu_{1}(r)\,=\,\frac{C_{3}}{r^{D-3}}+\frac{\left(g(R_0)-2\Lambda_{D}g'(R_0)\right)}
{(D-2)(D-1)}r^{2}
\label{lambda1}
\end{eqnarray}
Up to second order in $\alpha$ the equations are:
\begin{eqnarray}
(D-3)\mu_{2}(r)+r\mu_{2}'(r) +\frac{(g(R_0)-2 \Lambda_{D}g'(R_0))}{D-2}\left(g'(R_0)
-\frac{2D}{D-2}\Lambda_{D}g''(R_0)\right)r^2\,=\,0
\label{lambda_eqn2}
\end{eqnarray}
\begin{eqnarray}
&&\left[-C_{1} (D-3) r^{3-D}-\frac{2\Lambda_{D} r^2}{D-1}\right]\lambda_{2}(r)
+C_{2}(D-2)(D-1)\left(-\Lambda_{D}r^2+D-3\right)\mu_{2}(r)\nonumber\\
&-&\left(C_{1} r^{4-D}+r-\frac{r^3\Lambda_{D}}{D-1}\right)
\lambda_{2}'(r)-C_{3}C_{4} (D-2) (D-1) \left(-\Lambda_{D} r^2+D-3\right) r^{3-D}\nonumber\\
&-&C_{2}\left[(D-1)(C_{1}r^{3-D}+1)-\Lambda_{D} r^2\right]
\left[2\Lambda_{D}g'(R_0)^2 +g(R_0)\left(\frac{2D\Lambda_{D}g''(R_0)}{D-2}-g'(R_0)\right)
-\frac{4D\Lambda_{D}^{2}g'(R_0)g''(R_0)}{D-2}\right]r^{2}\nonumber\\
&-&C_{4}[C_{1}(D-1)r^{3-D}+2][2\Lambda_{D}g'(R_0)-g(R_0)]r^{2}\,=\,0\nonumber\\
&&
\label{mu_eqn2}
\end{eqnarray}
whose solutions are:
\begin{eqnarray}
\lambda_{2}(r)\,&=&\,C_{6}+\frac{C_{6} C_{1}+(C_{3}C_{4}-C_{2}C_{5})(D-2)(D-1)}{r^{D-3}}
\nonumber\\
&+&\left[-\frac{C_{6}\Lambda_{D}}{D-1}+\left(g(R_0)-2 \Lambda_{D}g'(R_0)\right)
\left(C_{4}+C_{2} g'(R_0)-\frac{2 C_{2} D\Lambda_{D}g''(R_0)}{D-2}\right)\right] r^2
\nonumber\\
&&
\label{lambda2}
\end{eqnarray}
\begin{eqnarray}
\mu_{2}(r)\,=\,\frac{C_{5}}{r^{D-3}}+\frac{\left(g(R_0)-2\Lambda_{D}g'(R_0)\right)
\left(2 D\Lambda_{D} g''(R_{0})-(D -2) g'(R_0)\right)}{(D -2)^2 (D -1)}r^{2}
\label{mu2}
\end{eqnarray}
Further orders in $\alpha^{3,4,...}$ can be obtained by inserting
previous results in the order $3,4,...$ ones to get
$\{\lambda_{3,4,...}(r),\mu_{3,4,...}(r) \}$ but of course the
corresponding equations become increasingly complicated.

Notice that from the obtained results up to second order in $\alpha$,
the corresponding metric has constant scalar curvature for any
value of the parameters $C_1, C_2, \dots, C_6$. As a matter of fact,
this metric is nothing but the standard Schwarzschild-$AdS$ geometry,
and can be easily rewritten in the usual form  by making a
trivial time reparametrization as follows:
\begin{eqnarray}
\overline{\lambda}(r)\,&\equiv&\,\lambda(r)\left[ -C_{2} (D ^2+3D-2)
+C_{4}\left(D^2-3D +2\right)\alpha +C_6\alpha^{2}+\Od(\alpha^{3})\right]\nonumber\\
\overline{\mu}(r)\,&\equiv&\,\mu(r)
\label{lambda_reparametrization}
\end{eqnarray}

Therefore, at least up to second order, the only static,
spherically symmetric solutions which are analytical
in $\alpha$ are the standard Schwarzschild-$AdS$ space-times.

On the other hand, taking the inverse point of view,
if we assume the solutions to be of the $AdS$ BH type at
any order in the $\alpha$ expansion we can write:
\begin{eqnarray}
\lambda(r)\,\equiv\,\mu(r)=\,1+\left(\frac{\overline{R}_{S}}{r}\right)^{D-3} + J r^2
\label{mu_lambda}
\end{eqnarray}
as solution for the Einstein equations \eqref{eqn_lambda} and \eqref{eqn_mu}
with the gravitational lagrangian \eqref{expansion_en_alpha_fR} and
\begin{eqnarray}
\overline{R}_{S}\,&=&\,R_{S}+\Sigma_{i=1}^{\infty} C_{i}\alpha^{i}\nonumber\\
J\,&=&\,-\frac{\Lambda_{D}}{(D-1)}+\Sigma_{i=1}^{\infty}J_{i}\alpha^{i}
\end{eqnarray}
where $R_S$ and $C_i$ are arbitrary constants and the $J_{i}$
coefficients can be determined from (\ref{root_R0}):
\begin{eqnarray}
R-(D-2)\Lambda_{D} +\alpha g(R)+2(D-1)J(1+\alpha g'(R))\,=\,0
\label{algebraic_eqn}
\end{eqnarray}
with $R\,=\,-D(D-1)J$. Expanding previous equation in powers of $\alpha$
it is possible to find a recurrence equation for the $J_i$ coefficients, namely
for the $J_l$ (with $l>0$) coefficient, we find:
\begin{eqnarray}
&&(2-D)(D-1)J_{l}+\sum_{i=0}^{l-1}\sum_{cond.1}\frac{1}{i_{1}!i_{2}!
\ldots i_{l-1}!}(J_{1})^{i_{1}}(J_{2})^{i_{2}} \ldots (J_{l-1})^{i_{l-1}}g^{(i)}(R_{0})
+\nonumber\\
&& 2(D-1)\sum_{k=0}^{l-1}J_{k}\sum_{i=0}^{l-k-1}\sum_{cond.2}
\frac{1}{i_{1}!i_{2}!\ldots i_{l-k-1}!}(J_{1})^{i_{1}}(J_{2})^{i_{2}}\ldots
(J_{l-k-1})^{i_{l-k-1}}g^{(i+1)}(R_{0}) \,=\,0
\end{eqnarray}
with $R_{0}=-D(D-1)J_{0}\,\equiv\,D\Lambda_{D}$, where the first sum
is done under the condition 1 given by:
\begin{eqnarray}
\sum_{m=1}^{l-1}i_{m}=i, \,\, i_{m}\,\in \, \Bbb{N}\cup\{0\}\,\,\; \mbox{and}\,\,\;
\sum_{m=1}^{l-1} m \,i_{m}=l-1
\end{eqnarray}
and the second one under the condition 2:
\begin{eqnarray}
\sum_{m=1}^{l-k-1}i_{m}=i, \,\, i_{m}\,\in \, \Bbb{N}\cup\{0\}\,\,\; \mbox{and}\,\,\;
\sum_{m=1}^{l-k-1} m \,i_{m}=l-k-1
\end{eqnarray}
For instance we have:
\begin{eqnarray}
J_{1}\,&=&\,
\frac{A(g\,;\,D,\,\Lambda_{D})}{(D-2)(D-1)}\nonumber\\
J_{2}\,&=&\,
- \frac{A(g\,;\,D,\,\Lambda_{D})[(D - 2) g'(R_{0})
- 2D\Lambda_{D}g''(R_{0})]}{(D - 2)^2 (D-1)}
\label{lambda&mu_expansions}
\end{eqnarray}
where $A(g\,;\,D,\,\Lambda_{D})\equiv g(R_0)-2\Lambda_{D} g'(R_0)$.

Now we can consider the possibility of  removing $\Lambda_{D}$ from
the action from the very beginning and still getting an $AdS$ BH
solution with an effective cosmological constant depending on
$g(R)$ and its derivatives evaluated at $R_{0}\equiv0 $. In this case the results,
order by order in $\alpha$ up to order $\alpha^2$, are:
\begin{eqnarray}
J_{0}(\Lambda_{D}=0)\,&=&\,0\nonumber\\
J_{1}(\Lambda_{D}=0)\,&=&\,\frac{g(0)}{(D -2) (D -1)}\nonumber\\
J_{2}(\Lambda_{D}=0)\,&=&\,-\frac{g(0) g'(0)}{(D -2) (D -1)}
\end{eqnarray}
As we see, in the context of $f(R)$ gravities, it is possible to have a BH in
an $AdS$ asymptotic space
even if the initial cosmological constant $\Lambda_D$ vanishes.

To end these two sections, we can summarize by saying that in the context
of $f(R)$ gravities the only  spherically symmetric and
static solutions of negative constant curvature are the standard BH in
$AdS$ space. The same result applies in the general case (without impossing
constant curvature) in perturbation theory up to second order. However,
the  possibility of having static and spherically symmetric solutions
with non constant curvature cannot be excluded in the case of
$f(R)$ functions which are not analytical in $\alpha$.

\section{Black-hole thermodynamics}

In order to consider the different thermodynamic quantities for
the $f(R)$ black-holes in $AdS$, we start from the temperature. In principle
there are two different ways of introducing this quantity for the
kind of solutions we are considering here. Firstly we can use the
definition coming from Euclidean quantum gravity \cite{HGG}. In this case one
introduces the Euclidean time $\tau=it$ and the Euclidean metric
$ds_E^2$ is defined as:
\begin{equation}
-\text{d}s_E^2=-\text{d}\sigma^2-r^2\text{d}\Omega^2_{D-2}
\end{equation}
where:
\begin{equation}
\text{d}\sigma^2=e^{-2\Phi(r)}A(r)\text{d}\tau^2+A^{-1}(r)\text{d}r^2.
\end{equation}
The metric corresponds only to the region $r>r_H$ where
$r_H$ is the outer horizon position with $A(r_H)=0$. Expanding
$\text{d}\sigma^2$ near $r_H$ we have:
\begin{equation}
\text{d}\sigma^2=e^{-2\Phi(r_H)}A'(r_H)\rho
\text{d}\tau^2+\frac{\text{d}\rho^2}{A'(r_H)\rho}
\end{equation}
where $\rho=r-r_H$. Now we introduce the new coordinates $\tilde R$ and
$\theta$ defined as:
\begin{eqnarray}
\theta=\frac{1}{2}e^{-\Phi(r_H)}A'(r_H)\tau
\nonumber\\
\tilde R=2\sqrt{\frac{\rho}{A'(r_H)}}
\end{eqnarray}
so that:
\begin{equation}
\text{d}\sigma^2=\tilde R^2\text{d}\theta^2+\text{d}R^2.
\end{equation}
According to the Euclidean quantum gravity prescription $\tau$
belongs to the interval defined by $0$ and $\beta_E=1/T_E$. On the
other hand, in  order to avoid conical singularities, $\theta$
must run between $0$ and $2\pi$. Thus it is found that
\begin{equation}
T_E=\frac{1}{4\pi} e^{ -\Phi(r_H) }    A'(r_H)
\end{equation}

Another possible definition of temperature was firstly proposed in
\cite{Hawking1974} stating that temperature can be given in terms of the
the horizon gravity $\mathcal{K}$ as :
\begin{eqnarray}
T_{\mathcal{K}}\equiv\frac{\mathcal{K}}{4\pi}
\end{eqnarray}
where $\mathcal{K}$ is given by:
\begin{eqnarray}
\mathcal{K}\,=\,\lim_{r\rightarrow
r_H}\frac{\partial_{r}g_{tt}}{\sqrt{|g_{tt}g_{rr}|}}.
\end{eqnarray}
Then it is straightforward to find:
\begin{eqnarray}
T_{\mathcal{K}}= T_E.
\end{eqnarray}
Therefore both definitions give the same result for this kind of
solution. Notice also that in any case the temperature depends
only on the behaviour of the metric near the horizon but it is
independent from the gravitational action. By this we mean that
different actions having the same solutions have also the same
temperature. This is not the case for other thermodynamic
quantities as we will see later. Taking into account the results
in previous sections and for simplicity we will concentrate only on
constant curvature $AdS$ BH solutions with $\Phi=0$ as a natural
choice and:
\begin{equation}
A(r)=1-\frac{R_S^{D-3}}{r^{D-3}}+\frac{r^2}{l^2}.
\end{equation}
Then, both definitions of temperature lead to:
\begin{equation}
\beta=1/T=\frac{4 \pi l^2 r_H}{(D-1)r_H^2+(D-3)l^2}.
\end{equation}
Notice that the temperature is a function of $r_H$ only, i.e. it
depends only on the BH size. In the limit $r_H$ going to zero the
temperature diverges as $T \sim 1/r_H$ and for $r_H$ going to
infinite $T$ grows linearly with $r_H$. Consequently $T$ has a
minimum at:
\begin{equation}
 r_{H0}=l\sqrt{\frac{D-3}{D-1}}
\end{equation}
  corresponding to a temperature:
\begin{equation}
T_0=\frac{\sqrt{(D-1)(D-3)}}{2 \pi l}
\end{equation}
The existence of this minimum was established in \cite{Hawking&Page}
for $D=4$ by Hawking and Page long time ago and it is well known.
More recently Witten extended this result to higher dimensions
\cite{Witten}. The minimun is important in order to set the regions
with different thermodynamic behaviors and stability properties.
 For $D=4$, an
exact solution can be found for $r_H$:
\begin{eqnarray}
r_{H}\,=\,l\frac{2^{1/3} \left(9 \frac{R_{S}}{l}+\sqrt{12+81 \frac{R_{S}^2}{l^2}}
\right)^{2/3}-(24)^{1/3}}{6^{2/3} \left(9\frac{R_S}{l}
+\sqrt{12+81\frac{R_S^2}{l^2}}\right)^{1/3}}
\end{eqnarray}
Thus, in the $R_S\ll l$ limit, we find $r_H=R_S$, whereas in the opposite
case $l\ll R_S$, we get $r_H=(l^2 R_S)^{1/3}$.
 For the particular
case $D=5$, $r_H$ can also be exactly found to be:
\begin{equation}
r_H^2=\frac{l^2}{2}\left(\sqrt{1+\frac{4R_S^2}{l^2}}-1\right)
\end{equation}
which goes to $R_S^2$ for $R_S\ll l$ and to $lR_S$ for $l\ll R_S$.
Notice that for any $T > T_0$, we have two
possible BH sizes: one corresponding to the small BH phase with
$r_H < r_{H0}$ and the
other corresponding to the large BH phase with $r_H > r_{H0}$.

In order to compute the remaining thermodynamic quantities, the Euclidean action
\begin{equation}
S_E=-\frac{1}{16 \pi G_D}\int \text{d}^{D}x\sqrt{g_E}\,(R+f(R))
\end{equation}
is considered. When the previous expression is evaluated on some
metric with a periodic Euclidean time with period $\beta$, it equals
$\beta$ times the free energy $F$ associated to this  metric.
Extending to the $f(R)$ theories, the computation by Hawking and
Page \cite{Hawking&Page}, generalized to higher dimensions by Witten
\cite{Witten}, we compute the difference of this action evaluated on
the BH and the $AdS$ metric which can be written as:
\begin{equation}
\Delta S_E=-\frac{R_0+f(R_0)}{16 \pi G_D}\Delta V
\end{equation}
where $R_0=-D(D-1)/l^2$ and $\Delta V$ is the volume difference
between both solutions which is given by:
\begin{equation}
\Delta V=\frac{\beta \mu_{D-2}}{2(D-1)}(l^2r^{D-3}_H-r^{D-1}_H)
\end{equation}
so that:
\begin{equation}
\Delta S_E=-\frac{(R_0+f(R_0))\beta \mu_{D-2}}{36 \pi (D-1)
G_D}(l^2r^{D-3}_H-r^{D-1}_H)=\beta F.
\end{equation}
Notice that from this expression it is straightforward to obtain the
free energy $F$. We see that provided $-(R_0+f(R_0))>0$, which is the usual
case in $EH$ gravity, we have
$F>0$ for $r_H<l$ and $F<0$ for $r_H>l$. The temperature corresponding to the
horizon radius $r_H=l$ will be denoted $T_1$ and it is given by:
\begin{equation}
T_1=\frac{D-2}{2\pi l}.
\end{equation}
Notice that for $D>2$ we have $T_0<T_1$.

On the other hand,  the total thermodynamical energy may now be obtained as:
\begin{equation}
   E=\frac{\partial \Delta S_E}{\partial \beta}=-\frac{(R_0+f(R_0))M l^2}{2(D-1)}
\end{equation}
where $M$ is the mass defined in \eqref{BHmass}. This is one of the possible
definitions for the BH energy for $f(R)$ theories, see for instance \cite{Multamaki2007}
for a more general discussion. For the $EH$
action we have $f(R)=-(D-2)\Lambda_D$ and then it is immediate to
find $E=M$. However this is not the case for general $f(R)$
actions. Notice, that positive energy in $AdS$ space-time requires
$R_0+f(R_0)<0$. Now the entropy $S$ can be obtained from the well-known
relation:
\begin{equation}
S=\beta E- \beta F.
\end{equation}
Then one gets:
\begin{equation}
S=-\frac{(R_0+f(R_0))l^2 A_{D-2}(r_H)}{8 (D-1)G_D}
\end{equation}
where $A_{D-2}(r_H)$ is the horizon area given by
$A_{D-2}(r_H)\equiv r_H^{D-2}\mu_{D-2}$. Notice that once again
positive entropy requires $R_0+f(R_0)<0$.
For the $EH$ action we
have $R_0+f(R_0)=-2(D-1)/l^2$ and then we get the famous
Hawking-Bekenstein result \cite{Bekenstein}
\begin{equation}
S=\frac{ A_{D-2}(r_H)}{4G_D}
\end{equation}
Finally we can compute the heat capacity $C$ which can be written
as:
\begin{equation}
C=\frac{\partial E}{\partial T}=\frac{\partial E}{\partial
r_H}\frac{\partial r_H}{\partial T}
\end{equation}
Then it is easy to find
\begin{equation}
C=\frac{-(R_0+f(R_0))(D-2)\mu_{D-2}r^{D-2}_Hl^2}{8G_D(D-1)}
\frac{(D-1)r^2_H+(D-3)l^2}{(D-1)r^2_H-(D-3)l^2}.
\end{equation}
For the particular case of the $EH$ action we find:
\begin{equation}
C=\frac{(D-2)\mu_{D-2}r^{D-2}_H}{4G_D}\frac{(D-1)r^2_H+(D-3)l^2}{(D-1)r^2_H-(D-3)l^2}.
\end{equation}
In the Schwarzschild limit $l$ going to infinity this formula
gives:
\begin{equation}
C=-\frac{(D-2)\mu_{D-2}r^{D-2}_H}{4G_D}< 0
\end{equation}
which is the negative well-known result for standard BH. In the
general case, assuming like in the $EH$ case $(R_0+f(R_0))<0$, we
find $C>0$ for $r_H > r_{H0}$ (the large BH region) and $C<0$ for
$r_H < r_{H0}$ (the small BH region). For $r_H \sim r_{H0}$ ($T$
close to $T_0$)   $C$ is divergent. Notice that in $EH$ gravity, $C<0$ necessarily
implies $F>0$ since $T_0<T_1$.

In any case, for $f(R)$ theories with $R_0+f(R_0)<0$, we have found
an scenario similar to the one described in
full detail by Hawking and Page in \cite{Hawking&Page} long time ago
for the $EH$ case.

For $T<T_0$, the only possible state of thermal equilibrium in  an
$AdS$ space is pure radiation with negative free energy and there is
no stable BH solutions. For $T>T_0$ we have two possible BH
solutions; the small (and light) BH and the large (heavy) BH. The
small one has negative heat capacity and positive free energy as the
standard Schwarzschild BH. Therefore it is unstable under Hawking
radiation decay. For the large BH we have two possibilities; if
$T_0<T<T_1$ then both, the heat capacity and the free energy are
positive and the BH will decay by tunneling into radiation, but if
$T>T_1$ then the heat capacity is still positive but the free energy
becomes negative. In this case the free energy of the heavy BH will
be less than that of pure radiation. Then pure radiation will tend
to tunnel or to collapse to the BH configuration in equilibrium with
thermal radiation.

In general $f(R)$ theories one could also in principle consider the
possibility of having $R_0+f(R_0)>0$. However in this case 
 the mass and the entropy would be negative and therefore
in such theories the $AdS$ BH solutions would be unphysical.
Therefore $R_0+f(R_0)<0$ can be regarded  as a necessary condition
for $f(R)$ theories in order to support $AdS$ BH solutions. Using
(\ref{root_R0}), this condition  implies $1+f'(R_0)>0$. This last
condition has a clear physical interpretation in $f(R)$ gravities
(see \cite{silvestri} and references therein). Indeed, it can be
interpreted as the condition for the effective Newton's constant
$G_{eff}=G_{D}/(1+f'(R_0))$ to be positive. It can also be
interpreted from the quantum point of view as the condition which
prevents the graviton from becoming a ghost.

\section{Particular examples}

In this section we will consider some particular $f(R)$ models in
order to calculate the heat capacity $C$ and the free energy $F$ as
the relevant thermodynamical quantities for local and global
stability of BH's. For these particular models, $R_0$ can be
calculated exactly by using \eqref{root_R0}. For the sake of
simplicity we will fix the $D$-dimensional Schwarzschild radius in
(\ref{BHmass}) as $R_S^{D-3}=2$. The models we have considered are:

\subsection{Model I: $f(R)\,=\,\alpha (-R)^{\beta}$ }
Substituting in \eqref{root_R0} for arbitrary dimension we get
\begin{eqnarray}
R \left[\left(1-\frac{2}{D}\right)-\alpha(-R)^{\beta-1}
\left(1-\frac{2}{D}\beta\right)\right] \,=\,0
\label{eqn_Model_I}
\end{eqnarray}

We will only consider non-vanishing curvature solutions, thus we find:
\begin{eqnarray}
R_{0}\,=\,-\left[\frac{2-D}{(2\beta-D)\alpha}\right]^{1/(\beta-1)}
\label{R0_Model_I}
\end{eqnarray}
Since $D$ is assumed to be larger than 2,
the condition $(2\beta-D)\alpha<0$ provides well defined scalar curvatures $R_{0}$. 
Two separated regions have thus to be studied: Region $1$ $\{\alpha<0,\,\beta>D/2\}$ 
and Region $2$ $\{\alpha>0,\, \beta<D/2\}$.
For this model we also get
\begin{eqnarray}
1+f'(R_{0})\,=\,\frac{D(\beta-1)}{2\beta-D}
\end{eqnarray}
Notice that in Region $1$, $1+f'(R_0)>0$ for $D>2$, since in this case $\beta>1$ 
is straightforwardly accomplished.
In Region $2$, we find that for $D>2$, the requirement $R_0+f(R_0)<0$, i.e. $1+f'(R_0)>0$, 
fixes $\beta<1$, since this is the most stringent constraint over the 
parameter $\beta$ in this region. Therefore the physical space of 
parameters in Region $2$ is restricted to be $\{\alpha>0,\,\beta<1\}$.

In Figs. 1-3 we plot the physical regions in the parameter space $(\alpha,\beta)$
corresponding to the different signs of $(C,F)$.
\subsection{Model II: $f(R)\,=\, -(-R)^{\alpha}\,\text{exp}(q/R)-R$}
In this case, a vanishing curvature solution appears
provided $\alpha>1$. In addition, we also have:
\begin{eqnarray}
R_{0}\,=\,\frac{2 q}{2\alpha-D}
\label{R0_model_II}
\end{eqnarray}
To get $R_{0}<0$ the condition $q(2\alpha-D)<0$ must hold and two separated
regions will be studied: Region $1$ $\{q>0,\, \alpha<D/2\}$ and Region $2$
$\{q<0,\,\alpha>D/2\}$.

In Figs. 4-6 we plot the regions in the parameter space $(\alpha,q)$
corresponding to the different signs of $(C,F)$.
\subsection{Model III: $f(R)\,=\, R\,(\text{log}\alpha R)^{q}-R$}
A vanishing curvature solution also appears in this model. The non trivial one
is given by
\begin{eqnarray}
R_{0}\,=\,\frac{1}{\alpha}\text{exp}\left(\frac{2 q}{D-2}\right)
\label{R0_model_III}
\end{eqnarray}
Since $R_{0}$ has to be
negative, $\alpha$ must be negative as well, accomplishing $\alpha R_{0}>0$ and
since $\alpha R$, and therefore $\alpha R_{0}$, has to be bigger than one to have
a positive number powered to $q$, what imposes $q>0$ as can be read from the argument of
the exponential in the previous equation. Therefore there exists a unique accessible
region for parameters in this model: $\alpha<0$ and $q>0$.

In Figs. 7-8 we plot the regions in the parameter space $(\alpha,q)$
corresponding to the different signs of $(C,F)$.
\subsection{Model IV: $f(R)\,=\,-\alpha\frac{c_1\left(\frac{R}{\alpha}\right)^{n}} {1+\beta\left(\frac{R}{\alpha}\right)^n}$}
This model has been proposed in \cite{Hu&Sawicki2007} as cosmologically viable. Throughout this section, we consider $n=1$ for this model. Hence imposing $f'(R_0)=\epsilon$ we get
\begin{eqnarray}
c_{1}\,=\,-\frac{(D - 2 (1 + \epsilon))^2}{D^2 \epsilon}
\label{epsilon}
\end{eqnarray}
hence a relation between $c_1$, $D$ and $\epsilon$ can be imposed and therefore this model would only depend on two parameters $\alpha$ and $\beta$.
A vanishing curvature solution also appears in this model and two non trivial
curvature solutions are given by:
\begin{eqnarray}
R_{0}^{\pm}\,=\,\frac{\alpha \left[(c_1-2) D+4\pm\sqrt{c_1} \sqrt{c_1 D^2-8 D+16}\right]}{2 \beta (D-2)}
\label{R0_model_IV}
\end{eqnarray}
The corresponding $1+f'(R_0)$ values for \eqref{R0_model_IV} are
\begin{eqnarray}
1+f'(R_{0}^{\pm})\,=\,1-\frac{4(D-2)^2}{\left(\sqrt{c_1 D^2-8D+16}\pm\sqrt{c_1} D\right)^2}
\label{1+f'(R0)_model_IV}
\end{eqnarray}
where $c_1>0$ and $c_1>(8D-16)/D^2$ are required for real $R_{0}$ solutions.  Since $1+f'(R_{0})>0$ is required, that
 means that $\text{sign}(R_{0}^{\pm})=\text{sign}(\alpha\beta)$. It can be shown that $1+f'(R_{0}^{-})$ is not
  positive for any allowed value of $c_1$ and therefore this curvature solution $R_{0}^{-}$ is excluded for our study.

$1+f'(R_{0}^+)>0$ only requires $c_1>0$ for dimension $D\geq4$ and
therefore $\epsilon<0$ is required according to \eqref{epsilon}.
Therefore only two accesible regions need to be studied:  Region $1$
$\{\alpha>0,\, \beta<0\}$ and Region $2$, $\{\alpha<0,\,\beta>0\}$.
\\

In Figs. 9-10 we plot the thermodynamical regions in the parameter space $(\alpha,\beta)$ for a chosen $\epsilon=-10^{-6}$. Note that $1+f'(R_{0}^{+})$ does depend neither on $\alpha$ nor on $\beta$ and that $R_{0}^{+}$ only depend on the quotient $\alpha/\beta$ for a fixed $c_1$.
\section{Conclusions}

In this work we have considered static spherically symmetric
solutions in $f(R)$ theories of gravity in arbitrary dimensions.
After discussing the constant curvature case (including charged
black-holes in 4 dimensions), we have studied the general case
without imposing, a priori, the condition of constant curvature. We
have performed a perturbative analysis around the $EH$ case which
makes possible to study those solutions which are regular in the
perturbative parameter $\alpha$. We have found explicit expressions
up to second order for the metric coefficients, which give rise to
constant curvature (Schwarzschild $AdS$) solutions as in the $EH$
case.

On the other hand, we have also calculated thermodynamical
quantities for the $AdS$ black holes and considered the issue of the
stability of this kind of solutions. We have found that the
condition for a $f(R)$ theory of gravity to support this kind of
black holes is given by $R_0+f(R_0)<0$ where $R_0$ is the constant
curvature of the $AdS$ space-time. This condition has been seen to
imply also that the effective Newton's constant is positive and that
the graviton does not become a ghost. For these $f(R)$ gravities the
qualitative thermodynamic behavior of the BH is the same as the one
found by Hawking and Page for the AdS BH but the value of some
thermodynamic magnitudes is different for different $f(R)$
gravities.

Finally we have considered several explicit examples of $f(R)$
functions and studied the parameter regions in which BH in such
theories are locally stable and globally preferred, finding the same
qualitative behaviour as in standard $EH$ gravity.
\\

{\bf Acknowledgements:} This work has been  supported by
Ministerio de Ciencia e Innovaci\'on (Spain) project numbers
FIS 2008-01323 and FPA
2008-00592, UCM-Santander PR34/07-15875 and UCM-BSCH GR58/08 910309.

\newpage

\begin{figure}[!hbp]
\subfigure[ \hspace{1ex} Model I, $D=4$, Region 1, $\alpha<0$, $\beta>2$.]{
\begin{overpic}[width=7.0cm]{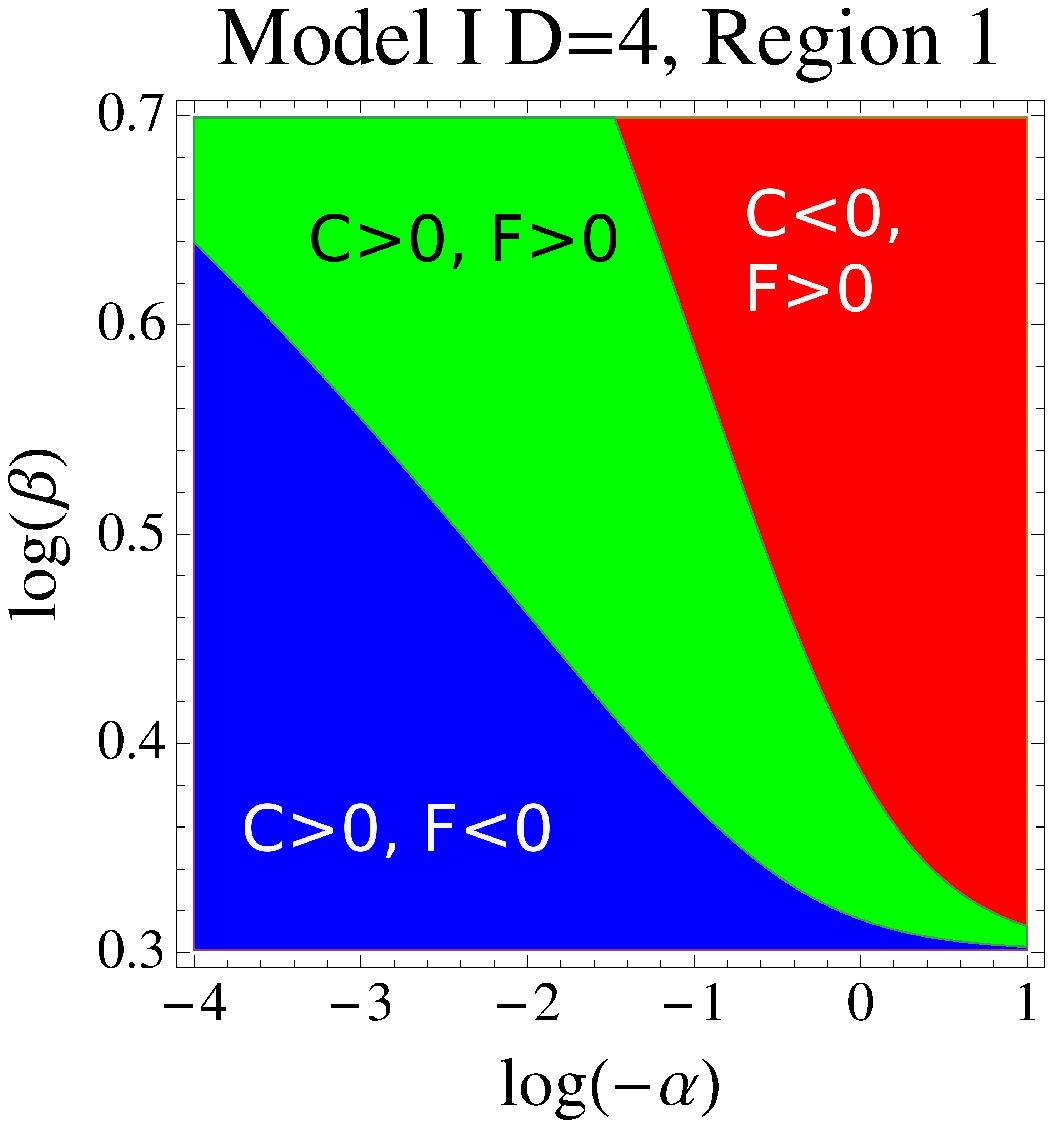}
\end{overpic}
}
\subfigure[ \hspace{1ex} Model I, $D=4$, Region 2, $\alpha>0$, $\beta<1$.]{
\begin{overpic}[width=7.70cm]{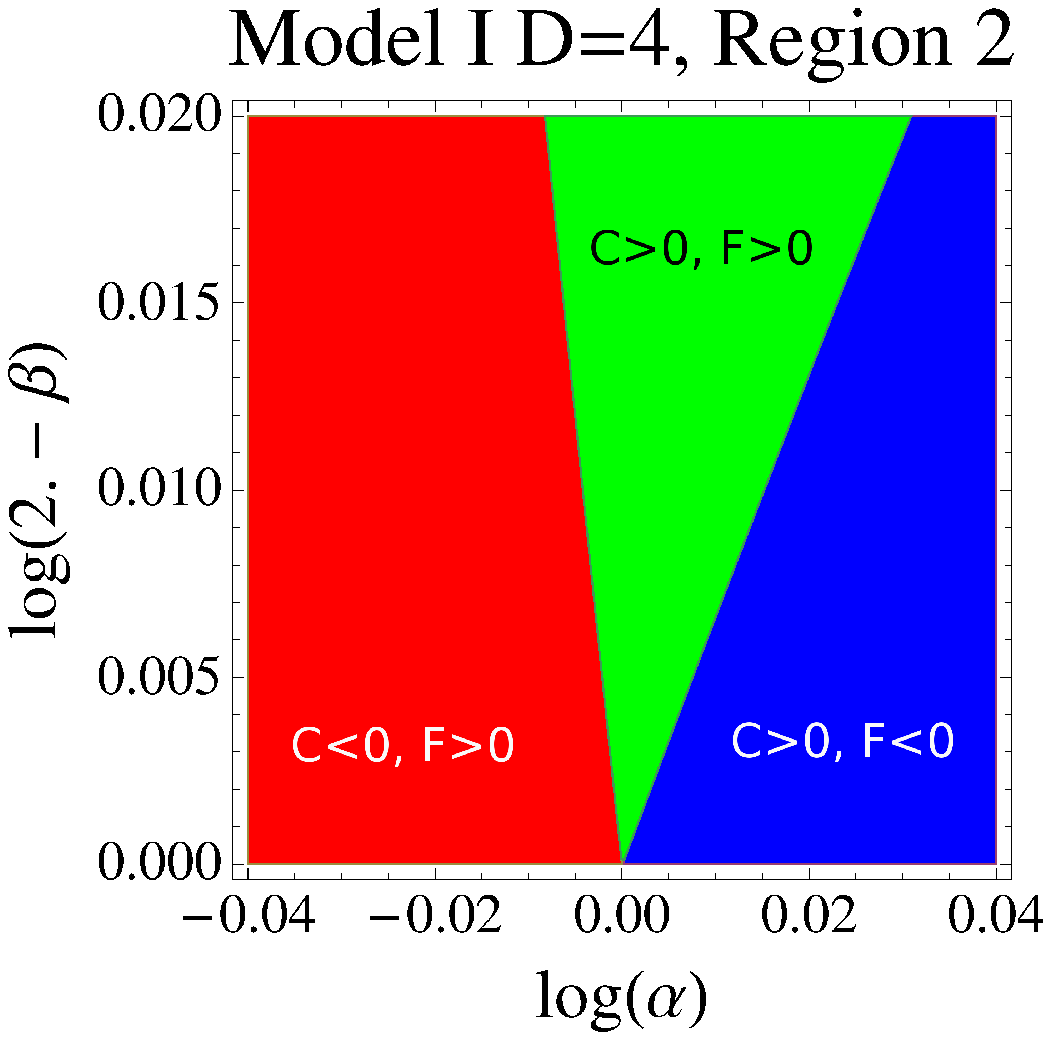}
\end{overpic}
}
\caption{Thermodynamical regions in the $(\alpha,\beta)$ plane for Model I in $D=4$.
Region 1(left),
Region 2 (right).}
\end{figure}
\begin{figure}[!hbp]
\subfigure[ \hspace{1ex} Model I, $D=5$, Region 1, $\alpha<0$, $\beta>2.5$.]{
\begin{overpic}[width=7.20cm]{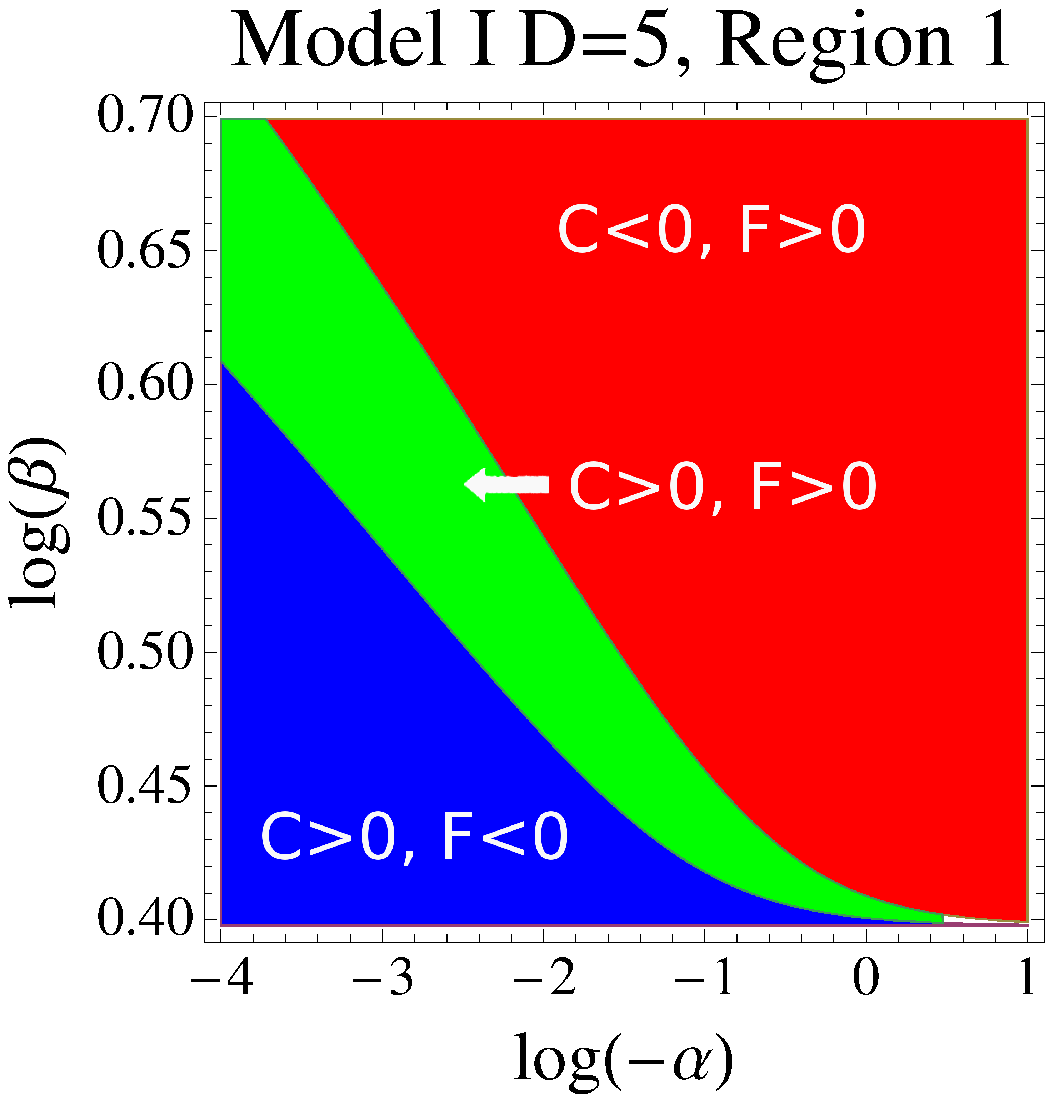}
\end{overpic}
}
\subfigure[ \hspace{1ex} Model I, $D=5$, Region 2, $\alpha>0$, $\beta<1$.]{
\begin{overpic}[width=7.70cm]{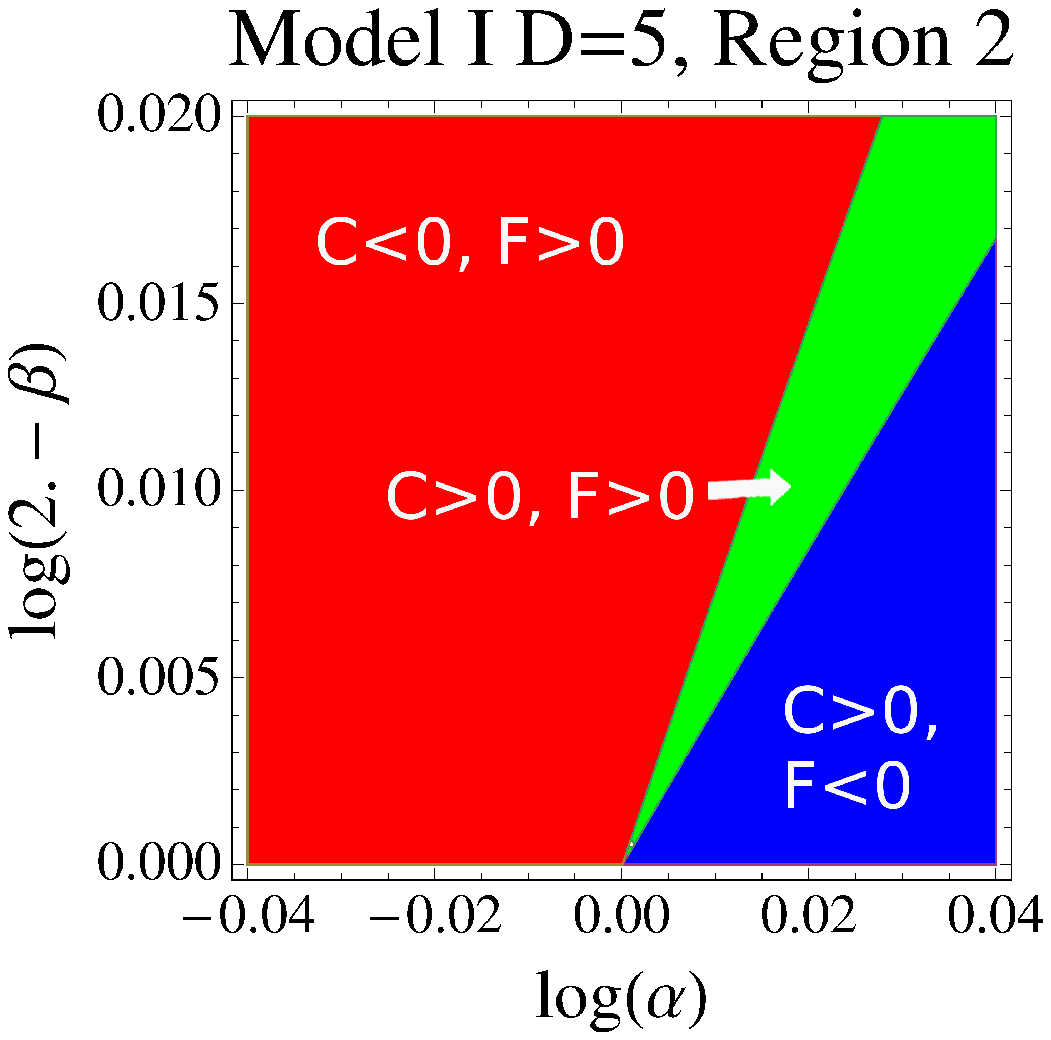}
\end{overpic}
}
\caption{Thermodynamical regions in the $(\alpha,\beta)$ plane for Model I in $D=5$.
Region 1(left),
Region 2 (right).}
\end{figure}
\begin{figure}[!hbp]
\subfigure[ \hspace{1ex} Model I, $D=10$, Region 1, $\alpha<0$, $\beta>5$.]{
\begin{overpic}[width=7.70cm]{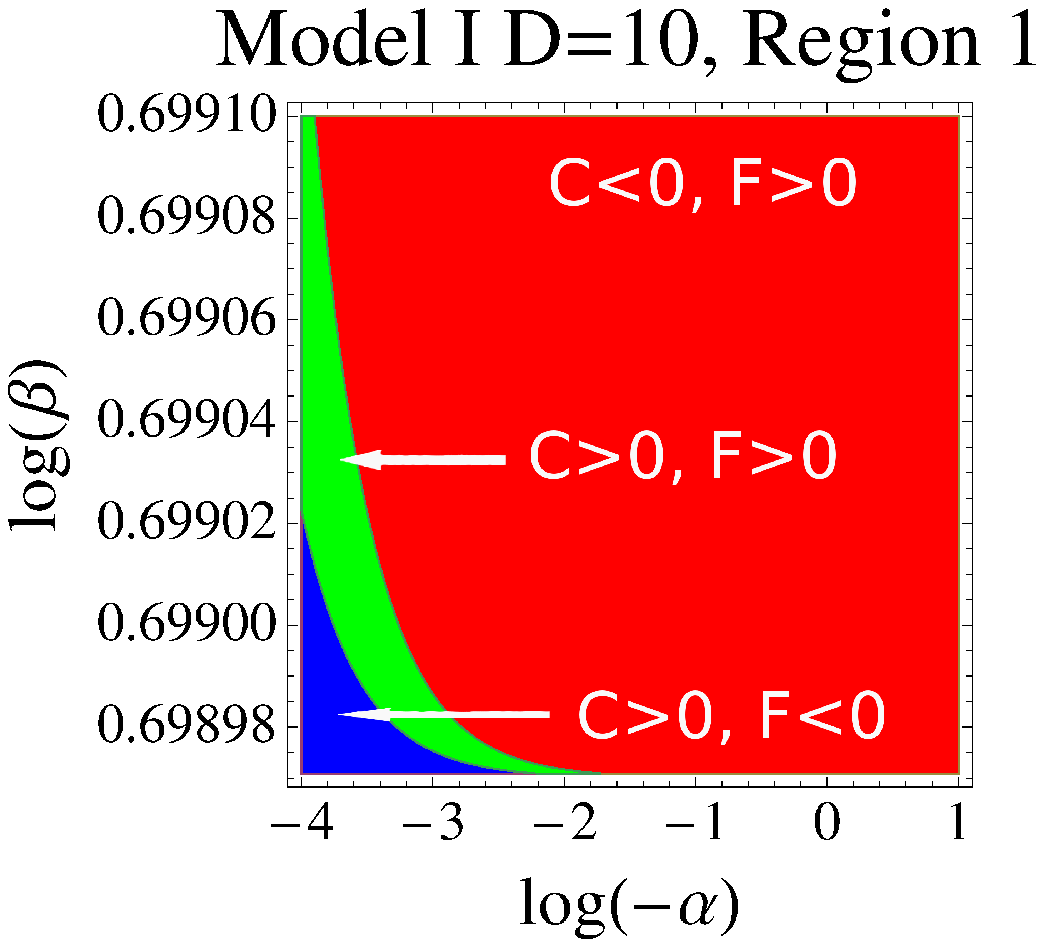}
\end{overpic}
}
\subfigure[ \hspace{1ex} Model I, $D=10$, Region 2, $\alpha>0$, $\beta<1$.]{
\begin{overpic}[width=7.0cm]{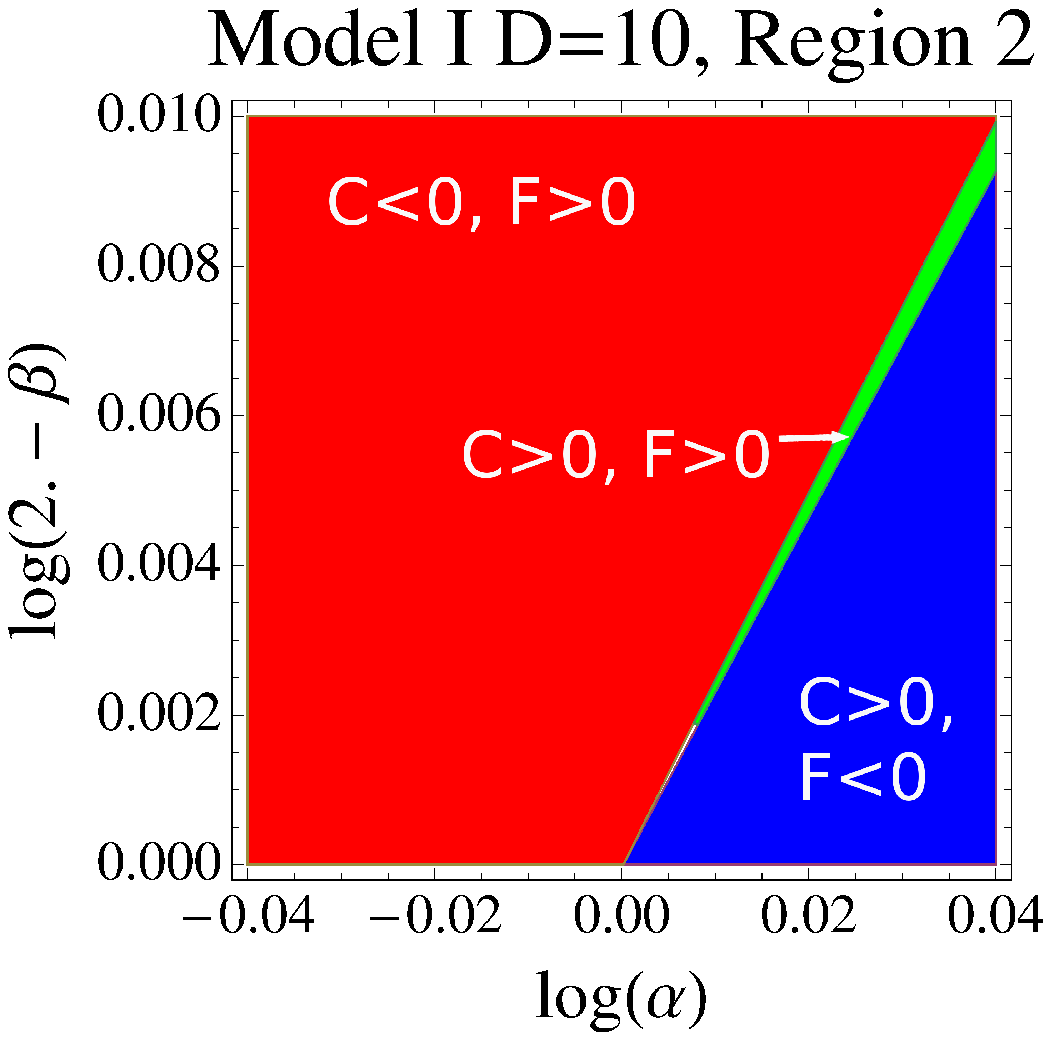}
\end{overpic}
}
\caption{Thermodynamical regions in the $(\alpha,\beta)$ plane for Model I in $D=10$.
Region 1(left),
Region 2 (right).}
\end{figure}
\begin{figure}[!hbp]
\subfigure[ \hspace{1ex} Model II, $D=4$, Region 1, $\alpha<2$, $q>0$.]{
\begin{overpic}[width=7.0cm]{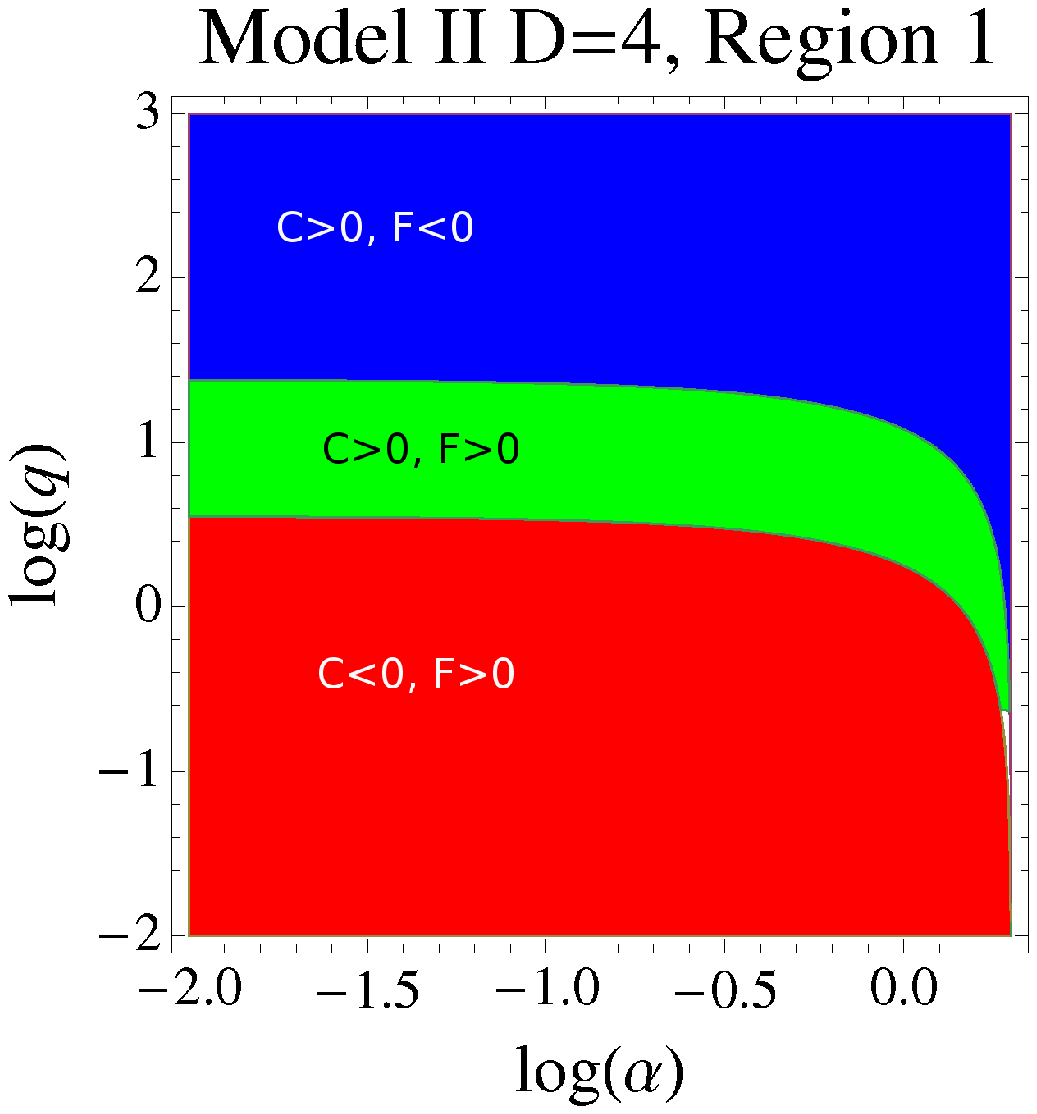}
\end{overpic}
}
\subfigure[ \hspace{1ex} Model II, $D=4$, Region 2, $\alpha>2$, $q<0$.]{
\begin{overpic}[width=7.0cm]{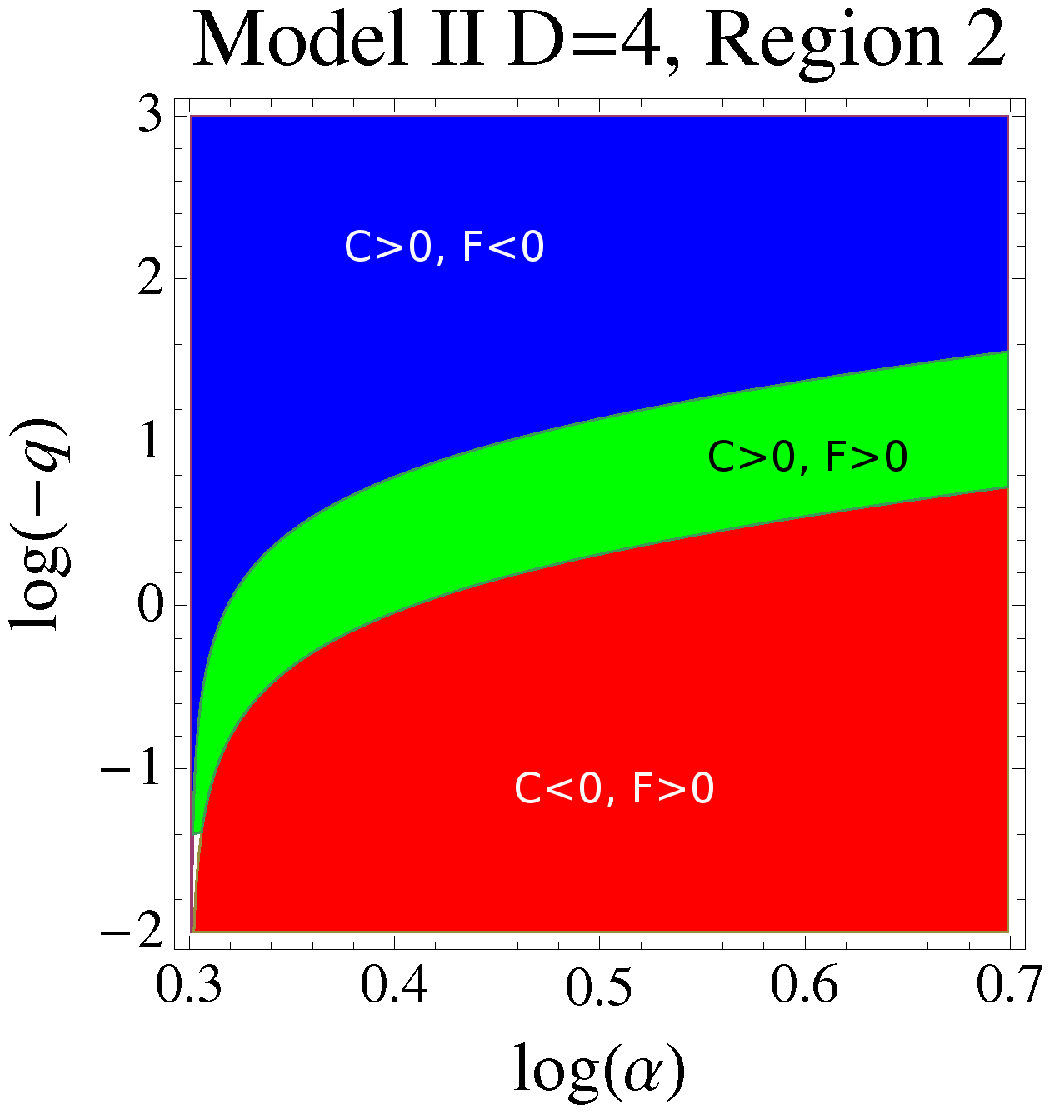}
\end{overpic}
}
\caption{Thermodynamical regions in the $(\alpha,q)$ plane for Model II in $D=4$.
Region 1(left),
Region 2 (right).}
\end{figure}

\begin{figure}[!hbp]
\centering
\subfigure[ \hspace{1ex} Model II, $D=5$, Region 1, $\alpha<2.5$, $q>0$.]{
\begin{overpic}[width=7.0cm]{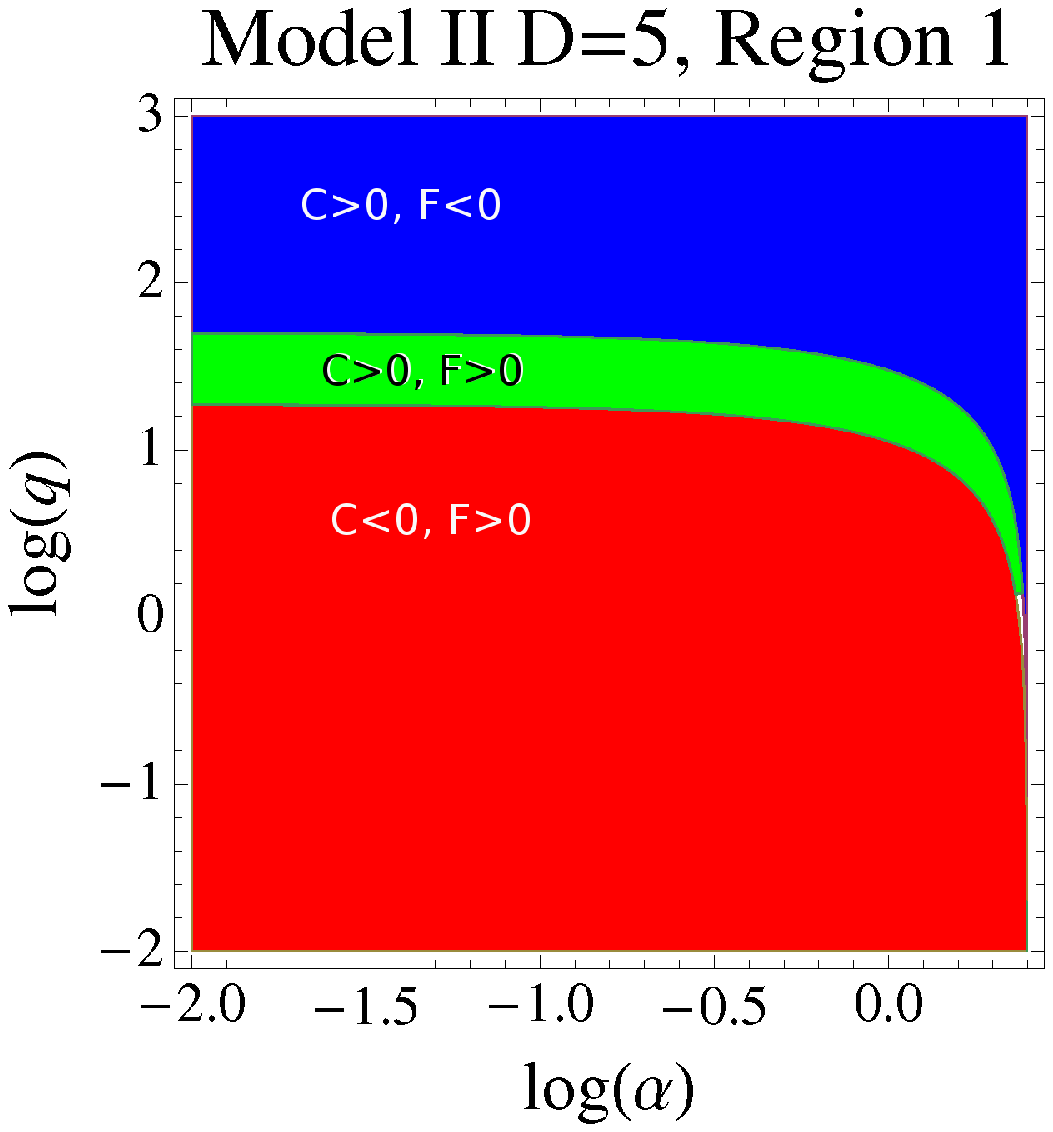}
\end{overpic}
}
\subfigure[ \hspace{1ex} Model II, $D=5$, Region 2, $\alpha>2.5$, $q<0$.]{
\begin{overpic}[width=7.0cm]{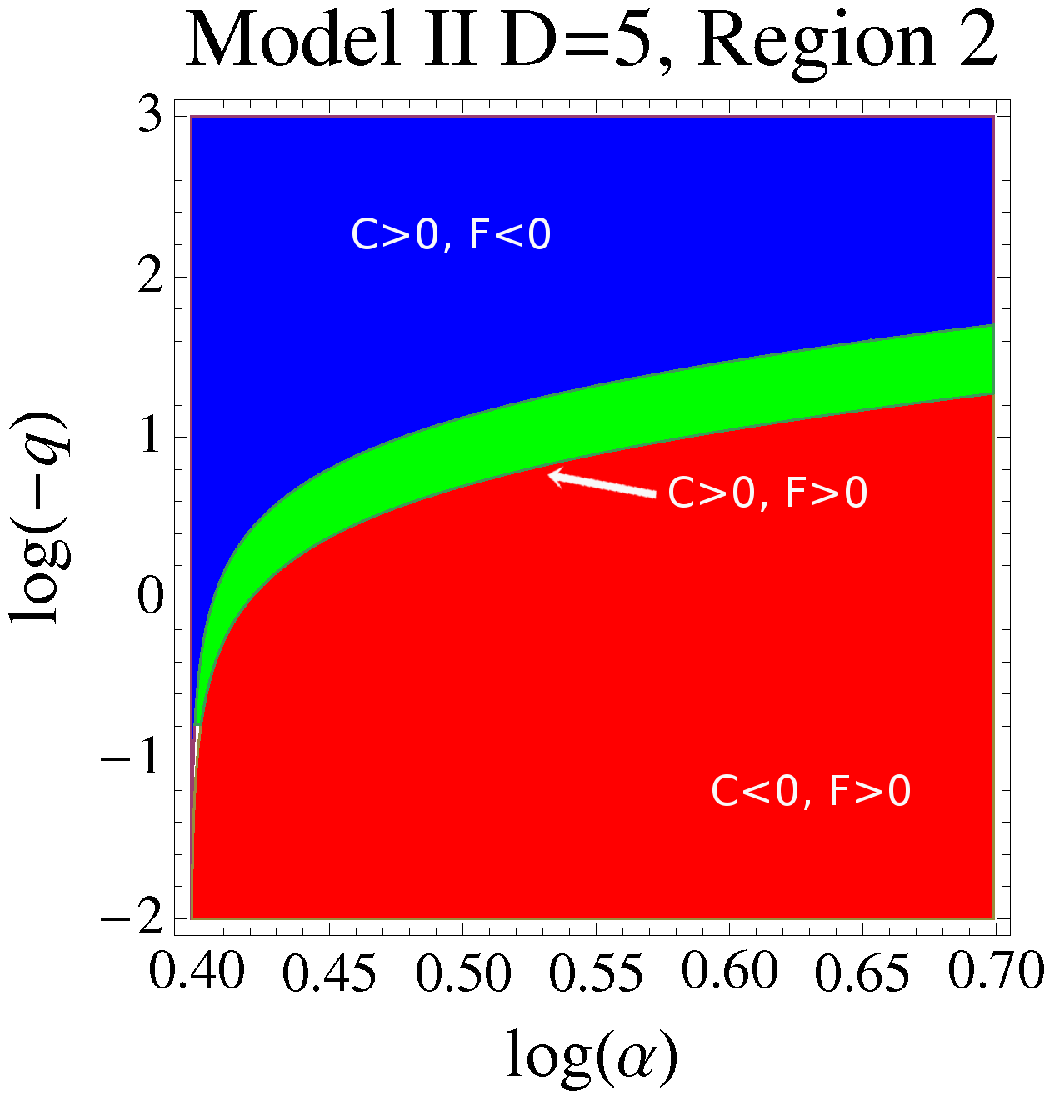}
\end{overpic}
}
\caption{Thermodynamical regions in the $(\alpha,q)$ plane for Model II in $D=5$.
Region 1(left), Region 2 (right).}
\end{figure}
\begin{figure}[!hbp]
\centering
\subfigure[ \hspace{1ex} Model II, $D=10$, Region 1, $\alpha<5$, $q>0$.]{
\begin{overpic}[width=7.0cm]{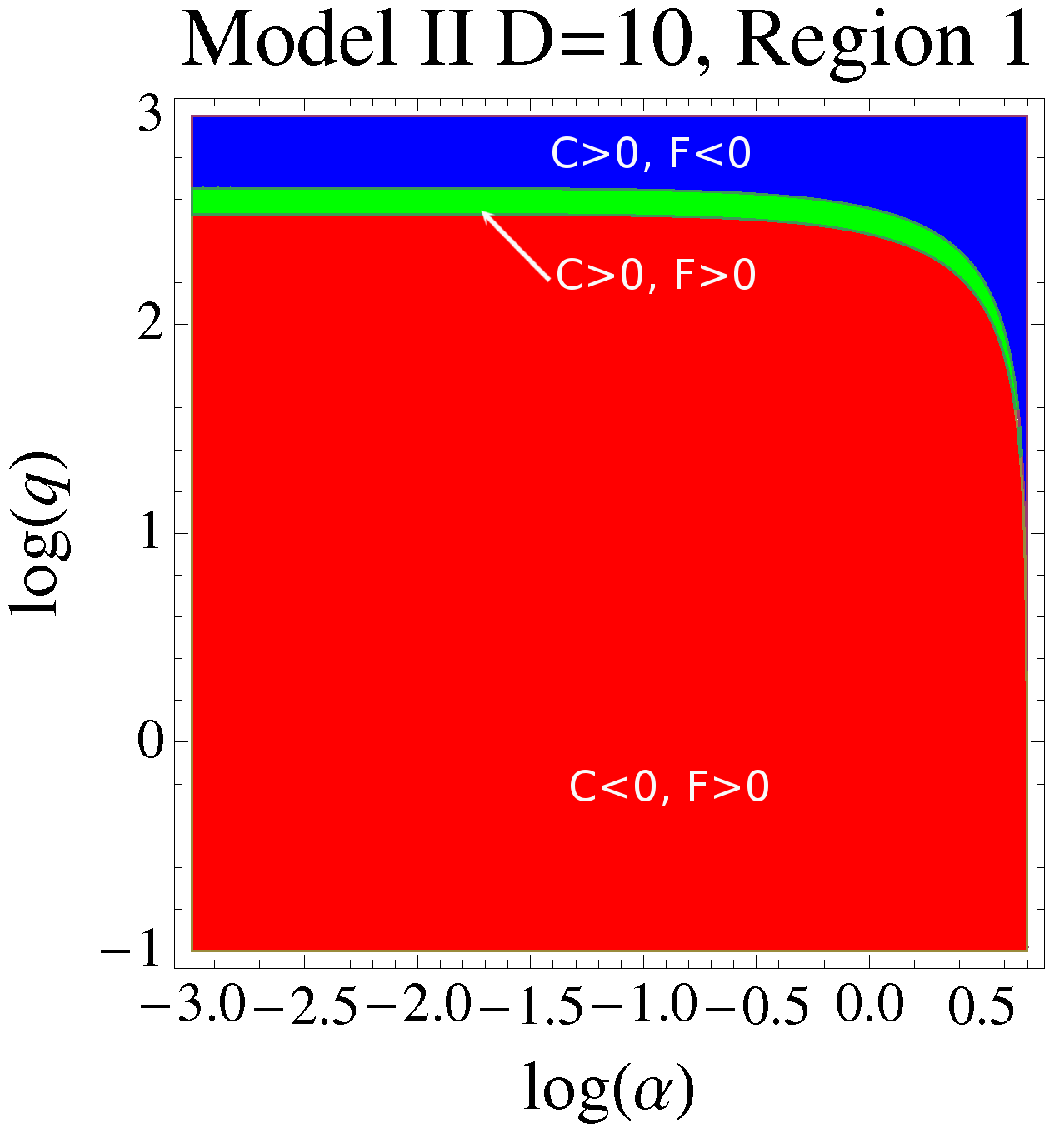}
\end{overpic}
}
\subfigure[ \hspace{1ex} Model II, $D=10$, Region 2, $\alpha>5$, $q<0$.]{
\begin{overpic}[width=7.0cm]{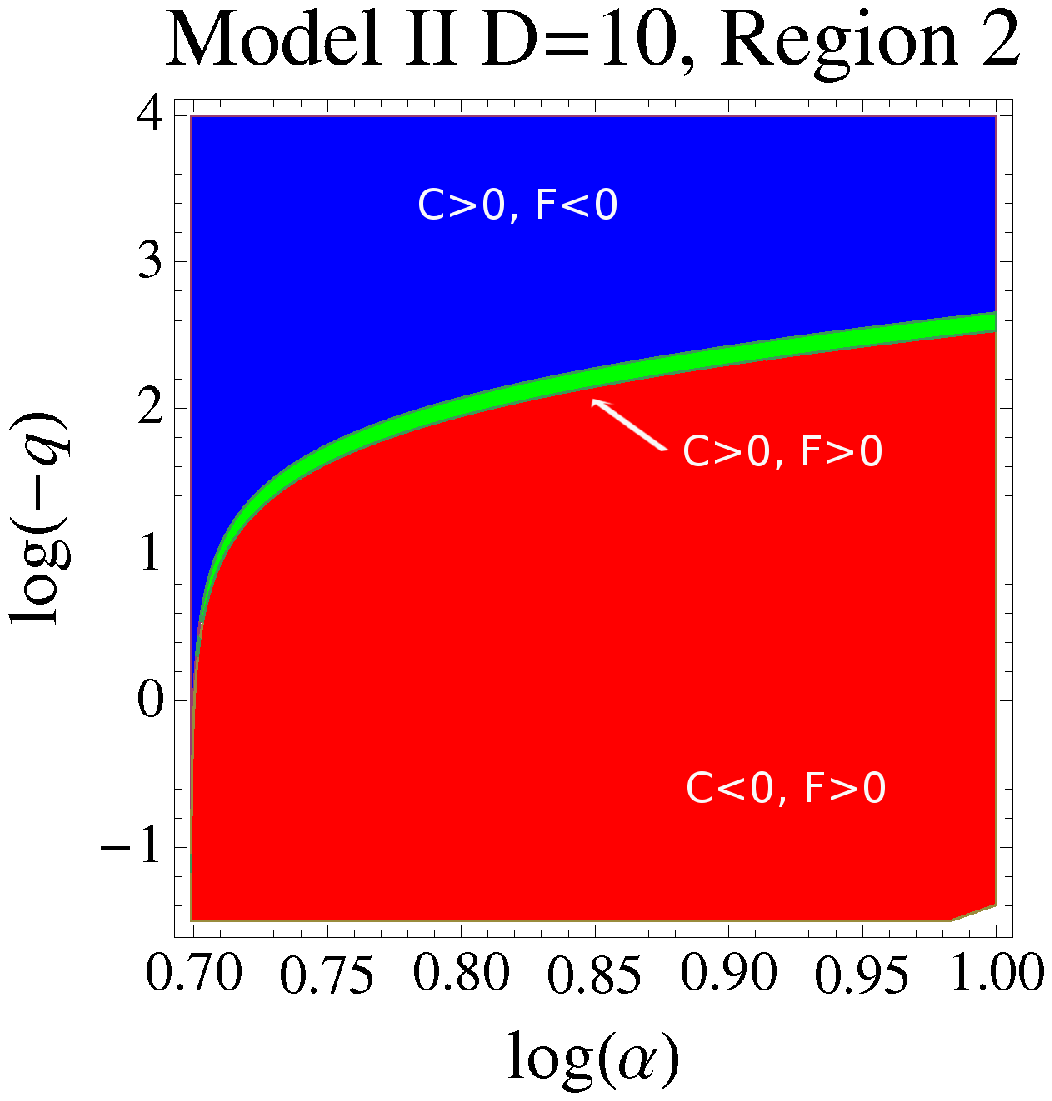}
\end{overpic}
}
\caption{Thermodynamical regions in the $(\alpha,q)$ plane for Model II in $D=10$.
Region 1(left), Region 2 (right).}
\end{figure}
\begin{figure}[!hbp]
\centering
\subfigure[ \hspace{1ex} Model III, $D=4$, $\alpha<0$, $q>0$.]{
\begin{overpic}[width=7.0cm]{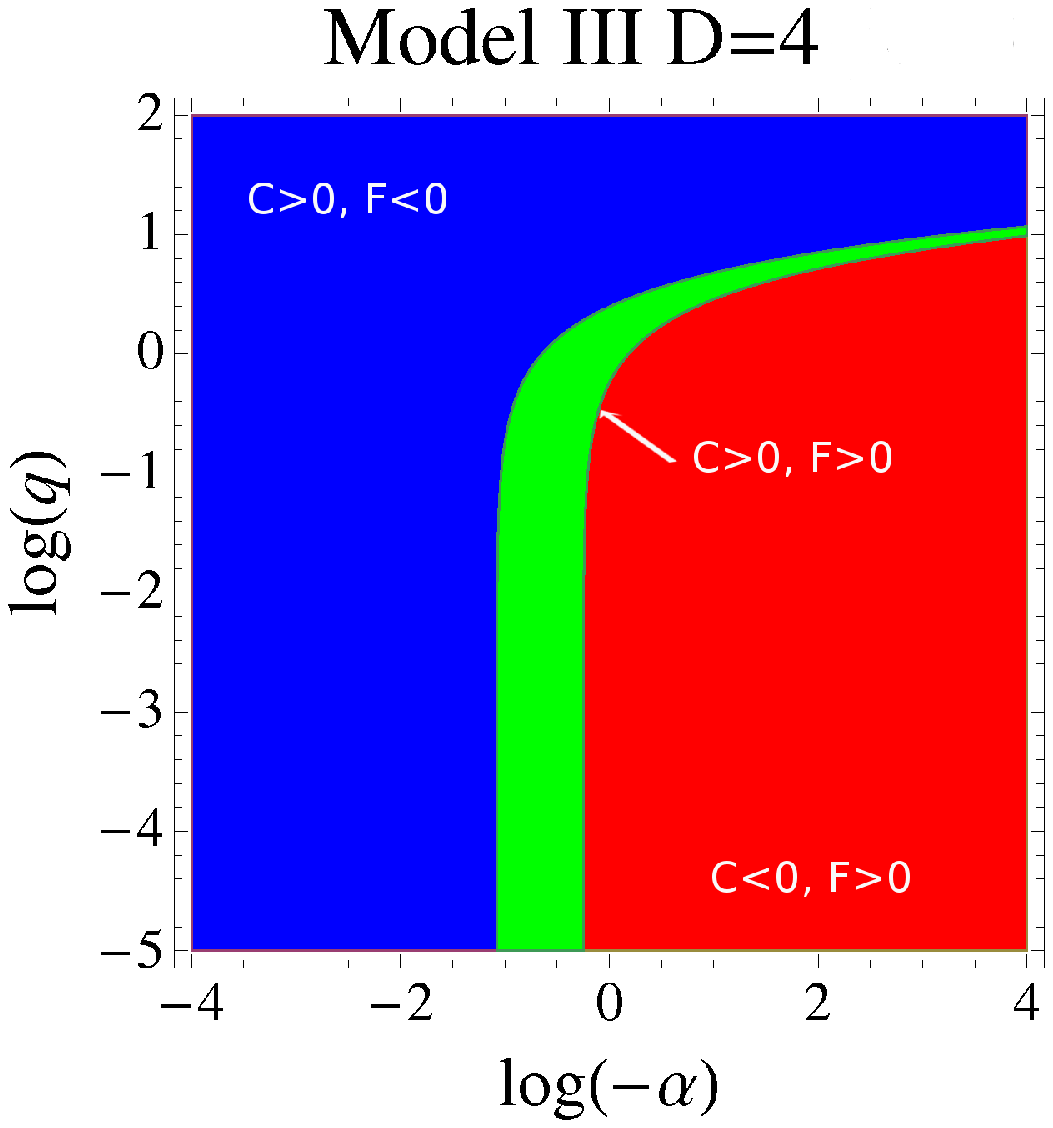}
\label{M2D4_g1}
\end{overpic}
}
\subfigure[ \hspace{1ex} Model III, $D=5$, $\alpha<0$, $q>0$.]{
\begin{overpic}[width=7.0cm]{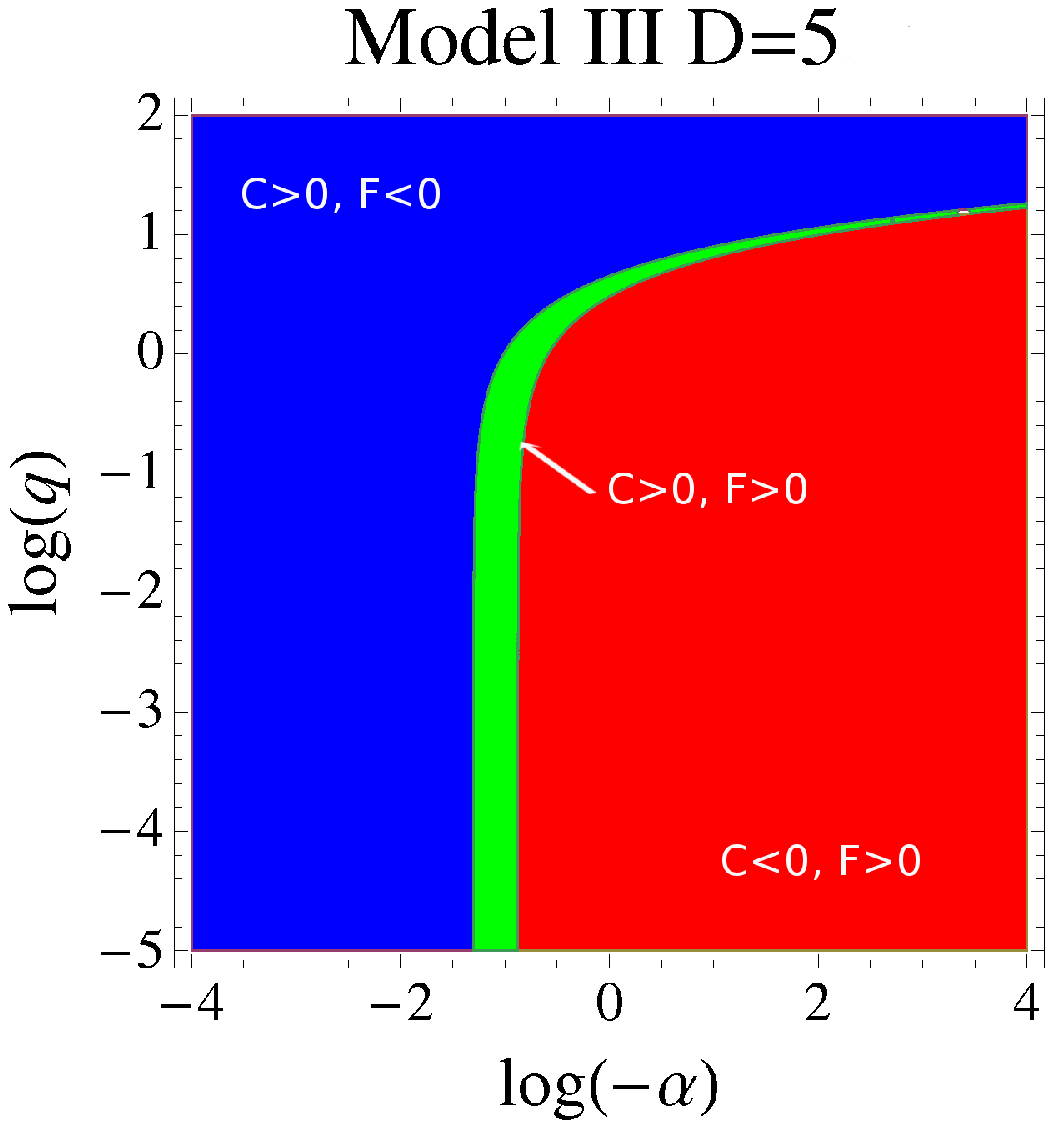}
\end{overpic}
}
\caption{Thermodynamical regions in the $(\alpha,q)$ plane for Model III in $D=4$ (left)
and $D=5$ (right).}
\end{figure}
\begin{figure}[!hbp]
\centering
\subfigure[ \hspace{1ex} Model III, $D=10$, $\alpha<0$, $q>0$.]{
\begin{overpic}[width=7.0cm]{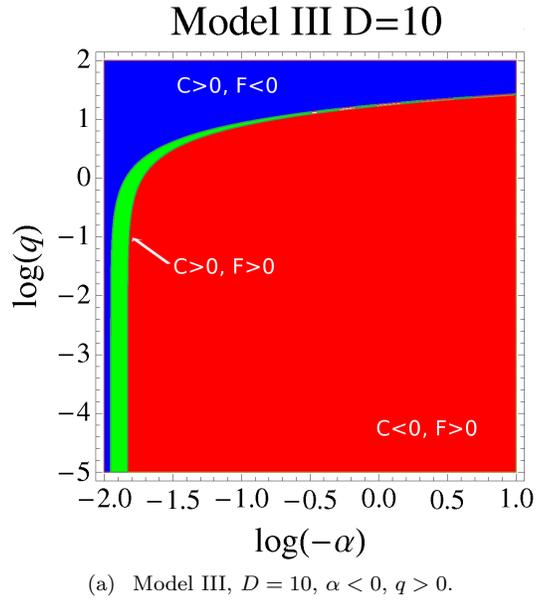}
\end{overpic}
}
\caption{Thermodynamical regions in the $(\alpha,q)$ plane for Model III in $D=10$.}
\end{figure}
\begin{figure*}[!hbp]
\subfigure{
\begin{overpic}[width=7.0cm]{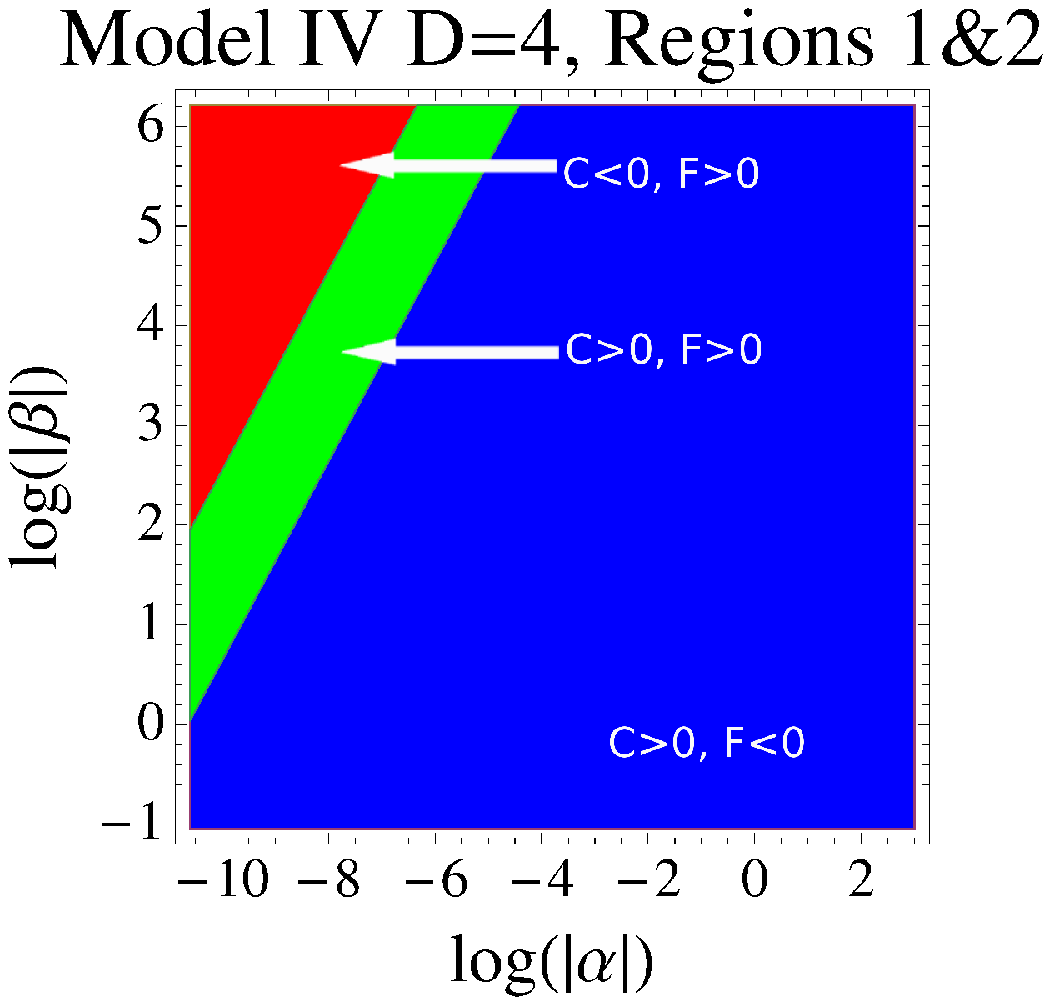}
\end{overpic}
}
\subfigure{
\begin{overpic}[width=7.0cm]{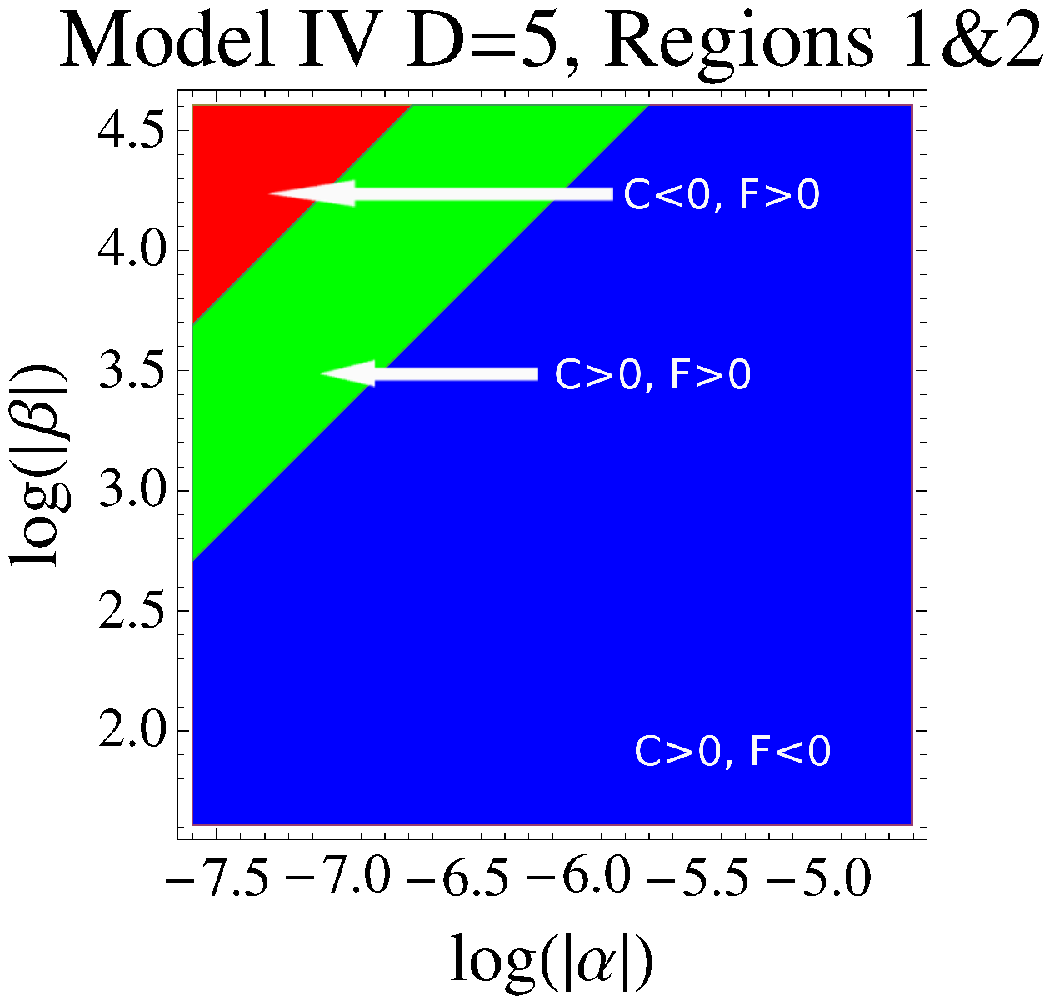}
\end{overpic}
}
\caption{Thermodynamical regions in the $(|\alpha|,|\beta|)$ plane
 for Model IV in $D=4$  (left) and  $D=5$ (right).}
\end{figure*}
\begin{figure*}[!hbp]
\subfigure{
\begin{overpic}[width=7.0cm]{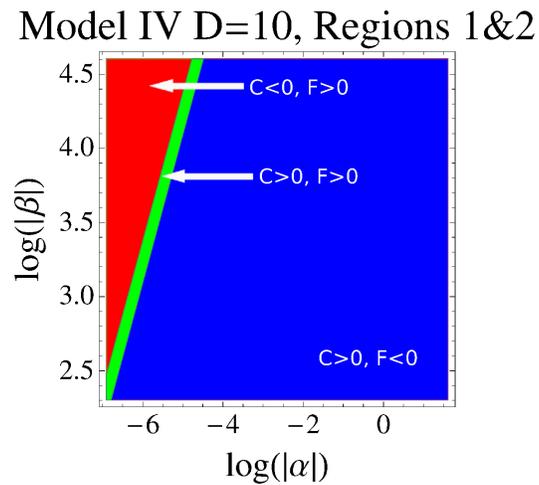}
\end{overpic}
}
\caption{Thermodynamical regions in the $(|\alpha|,|\beta|)$ plane
 for Model IV in $D=10$.}
\end{figure*}
%

\end{document}